\documentclass[1p]{elsarticle}

\usepackage{graphicx}
\usepackage{subfig}

\usepackage{amssymb}
\usepackage{gensymb}

\usepackage{lineno}

\biboptions{comma,square,sort&compress}

\journal{Nuclear Instruments and Methods A}

\usepackage{cleveref}
\crefname{figure}{Fig.\@}{Figs.\@}
\crefname{equation}{Eq.\@}{Eqs.\@}
\crefname{section}{Sec.\@}{Secs.\@}
\crefname{table}{Table}{Tables}

\begin{document}


\begin{frontmatter}

\title{The OLYMPUS Experiment\\
\vskip0.2cm
{\normalsize The OLYMPUS Collaboration}}

\author[inst9]{R.~Milner}
\author[inst9]{D.K.~Hasell\corref{cor1}}
\cortext[cor1]{Corresponding Author}
\ead{hasell@mit.edu}
\author[inst4]{M.~Kohl}
\author[inst2]{U.~Schneekloth}
\author[inst13]{N.~Akopov}
\author[inst1]{R.~Alarcon}
\author[inst10]{V.A.~Andreev}
\author[inst4]{O.~Ates}
\author[inst13]{A.~Avetisyan}
\author[inst3]{D.~Bayadilov}
\author[inst3]{R.~Beck}
\author[inst10]{S.~Belostotski}
\author[inst9]{J.C.~Bernauer}
\author[inst9]{J.~Bessuille}
\author[inst2]{F.~Brinker}
\author[inst9]{B.~Buck}
\author[inst12]{J.R.~Calarco}
\author[inst6]{V.~Carassiti}
\author[inst7]{E.~Cisbani}
\author[inst6]{G.~Ciullo}
\author[inst6]{M.~Contalbrigo}
\author[inst2]{N.~D'Ascenzo}
\author[inst5]{R.~De Leo}
\author[inst4]{J.~Diefenbach\fnref{fn1}}
\fntext[fn1]{Currently with Johannes Gutenberg-Universit\"at, Mainz, Germany}
\author[inst9]{T.W.~Donnelly}
\author[inst9]{K.~Dow}
\author[inst13]{G.~Elbakian}
\author[inst3]{D.~Eversheim}
\author[inst7]{S.~Frullani}
\author[inst3]{Ch.~Funke}
\author[inst10]{G.~Gavrilov}
\author[inst8]{B.~Gl\"aser}
\author[inst2]{N.~G\"orrissen}
\author[inst2]{J.~Hauschildt }
\author[inst9]{B.S.~Henderson}
\author[inst3]{Ph.~Hoffmeister}
\author[inst2]{Y.~Holler}
\author[inst1]{L.D.~Ice}
\author[inst10]{A.~Izotov}
\author[inst11]{R. Kaiser}
\author[inst13]{G.~Karyan}
\author[inst9]{J.~Kelsey}
\author[inst8]{D.~Khaneft}
\author[inst3]{P.~Klassen}
\author[inst10]{A.~Kiselev\fnref{fn4}}
\fntext[fn4]{Currently with Brookhaven National Laboratory, Upton, NY, USA}
\author[inst10]{A.~Krivshich}
\author[inst11]{I.~Lehmann\fnref{fn6}}
\fntext[fn6]{Also with the Facility for Antiproton and Ion Research, Darmstadt, Germany}
\author[inst6]{P.~Lenisa}
\author[inst2]{D.~Lenz}
\author[inst11]{S.~Lumsden}
\author[inst8]{Y.~Ma\fnref{fn2}}
\fntext[fn2]{Currently with RIKEN, Nishina Center, Advanced Meson Science Laboratory, Japan}
\author[inst8]{F.~Maas}
\author[inst13]{H.~Marukyan}
\author[inst10]{O.~Miklukho}
\author[inst6,inst13]{A.~Movsisyan}
\author[inst11]{M.~Murray}
\author[inst10]{Y.~Naryshkin}
\author[inst9]{C.~O'Connor}
\author[inst8]{R.~Perez Benito}
\author[inst5]{R.~Perrino}
\author[inst9]{R.P.~Redwine}
\author[inst8]{D.~Rodr\'iguez~Pi\~neiro}
\author[inst11]{G.~Rosner\fnref{fn6}}
\author[inst9]{R.L.~Russell}
\author[inst9]{A.~Schmidt}
\author[inst11]{B.~Seitz}
\author[inst6]{M.~Statera}
\author[inst3]{A.~Thiel}
\author[inst13]{H.~Vardanyan}
\author[inst10]{D.~Veretennikov}
\author[inst9]{C.~Vidal}
\author[inst9]{A.~Winnebeck\fnref{fn3}}
\fntext[fn3]{Currently with Varian Medical Systems, Bergisch Gladbach, Germany}
\author[inst13]{V.~Yeganov }

\address[inst9]{Massachusetts Institute of Technology, Cambridge, MA, USA}
\address[inst4]{Hampton University, Hampton, VA, USA}
\address[inst2]{Deutsches Elektronen-Synchrotron DESY, Hamburg, Germany}
\address[inst13]{Alikhanyan National Science Laboratory (Yerevan Physics
Institute), Yerevan, Armenia}
\address[inst1]{Arizona State University, Tempe, AZ, USA}
\address[inst10]{Petersburg Nuclear Physics Institute, Gatchina, Russia}
\address[inst3]{Friedrich Wilhelms Universit\"at, Bonn, Germany}
\address[inst12]{University of New Hampshire, Durham, NH, USA}
\address[inst6]{Universit{\`a} di Ferrara and Istituto Nazionale di Fisica Nucleare, Ferrara, Italy}
\address[inst7]{Istituto Superiore di Sanit\`{a} and Istituto Nazionale di Fisica Nucleare, Rome, Italy}
\address[inst5]{Istituto Nazionale di Fisica Nucleare, Bari, Italy}
\address[inst8]{Johannes Gutenberg-Universit\"at, Mainz, Germany}
\address[inst11]{University of Glasgow, Glasgow, United Kingdom}

\begin{abstract}
  The OLYMPUS experiment was designed to measure the ratio between the
  positron-proton and electron-proton elastic scattering cross
  sections, with the goal of determining the contribution of
  two-photon exchange to the elastic cross section. Two-photon
  exchange might resolve the discrepancy between measurements of the
  proton form factor ratio, $\mu_p G^p_E/G^p_M$, made using
  polarization techniques and those made in unpolarized
  experiments. OLYMPUS operated on the DORIS storage ring at DESY,
  alternating between 2.01~GeV electron and positron beams incident on
  an internal hydrogen gas target. The experiment used a toroidal
  magnetic spectrometer instrumented with drift chambers and
  time-of-flight detectors to measure rates for elastic scattering
  over the polar angular range of approximately
  $25\degree$--$75\degree$.  Symmetric M{\o}ller/Bhabha calorimeters
  at $1.29\degree$ and telescopes of GEM and MWPC detectors at
  $12\degree$ served as luminosity monitors. A total luminosity of
  approximately 4.5~fb$^{-1}$ was collected over two running periods
  in 2012. This paper provides details on the accelerator, target,
  detectors, and operation of the experiment.
\end{abstract}


\end{frontmatter}


\section{Introduction}
\label{sec:intro}

Electron scattering has long been an important tool for studying the
structure of nucleons. The strength of the technique lies in the
predominantly electromagnetic nature of the interaction. The electron
is, to the best of our knowledge, a point-particle, and its
interaction is well described by quantum electrodynamics. The
interaction is mediated by a virtual photon, whose momentum transfer
sets a size scale for the structures that are probed in the scattering
reaction. A low-momentum virtual photon can only ``see'' the size of
the nucleon, but by increasing the momentum transfer, the photon is
sensitive to the nucleon's internal distribution of charge and
magnetism, parameterized by form factors $G_E$ and $G_M$. At even
higher momentum transfers, deep inelastic scattering reveals the
distributions of the quarks and gluons, which are ultimately
responsible for the observed form factors. The synthesis of data at
all different momentum scales can verify and guide our theoretical
understanding of the nucleon.

\begin{figure}[htbp]
\centering
\includegraphics[width=\columnwidth]{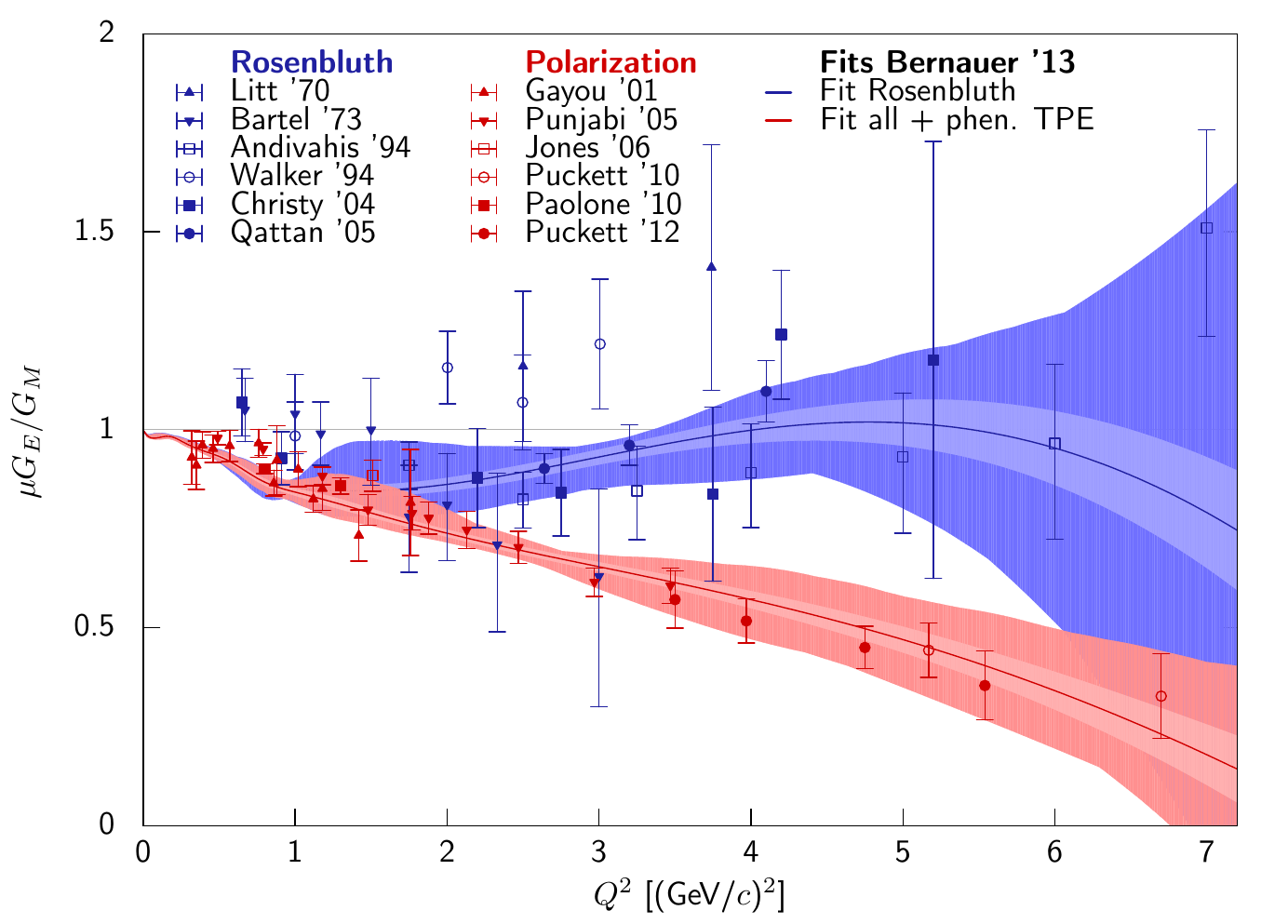}
\caption{The ratio of proton form factors, $\mu_p G^p_E / G^p_M$, as a
  function of $Q^2$ from (unpolarized) Rosenbluth
  measurements~\citep{Litt:1969my, Bartel:1973rf, Andivahis:1994rq,
    Walker:1993vj, Christy:2004rc, Qattan:2004ht} are inconsistent
  with recent data collected using polarization
  techniques~\citep{Gayou:2001qt, Punjabi:2005wq, Jones:2006kf,
    Puckett:2010ac, Paolone:2010qc,Puckett:2011xg}.  Also shown are
  the ratios from fits~\citep{Bernauer:2013tpr} of the form factors to
  the world dataset. The light shaded bands show statistical
  uncertainty and the dark shaded bands show model uncertainty added
  linearly.}
\label{fig:GEGMRatio}
\end{figure}

Polarized beams and targets offer another window into the structure of
nucleons. Recently, measurements of the electric to magnetic form
factor ratio of the proton, $\mu_p G^p_E/G^p_M$, using polarization
techniques~\citep{Milbrath:1997de, Pospischil:2001pp, Hu:2006fy,
  MacLachlan:2006vw, Crawford:2006rz, Ron:2011rd, Zhan:2011ji,
  Gayou:2001qt, Punjabi:2005wq, Jones:2006kf, Puckett:2010ac,
  Paolone:2010qc,Puckett:2011xg} have shown a dramatic discrepancy at
high four-momentum transfer, $Q^2$, in comparison with the ratio
obtained using the traditional Rosenbluth technique in unpolarized
cross section measurements~\citep{Litt:1969my, Bartel:1973rf,
  Andivahis:1994rq, Walker:1993vj, Christy:2004rc, Qattan:2004ht},
highlighted in Fig.~\ref{fig:GEGMRatio} by a selection of data sets.
This discrepancy might arise from a significant contribution to the
elastic electron-proton cross section from hard two-photon
exchange~\citep{ Guichon:2003qm, Blunden:2003sp, Chen:2004tw,
  Afanasev:2005mp, Blunden:2005ew, Kondratyuk:2005kk}, a process that
is neglected in the standard radiative corrections procedures. Since
there is no theoretical consensus on the size of this
contribution~\citep{Guichon:2003qm, Blunden:2003sp, Chen:2004tw,
  Afanasev:2005mp, Blunden:2005ew, Kondratyuk:2005kk, Borisyuk:2006fh,
  TomasiGustafsson:2009pw, TomasiGustafsson:2004ms,
  Bystritskiy:2007hw,Gorchtein:2006mq}, definitive measurements are
needed to determine if two-photon exchange resolves the form factor
discrepancy.

\begin{figure}[htbp]
\centering
\includegraphics[width=\columnwidth]{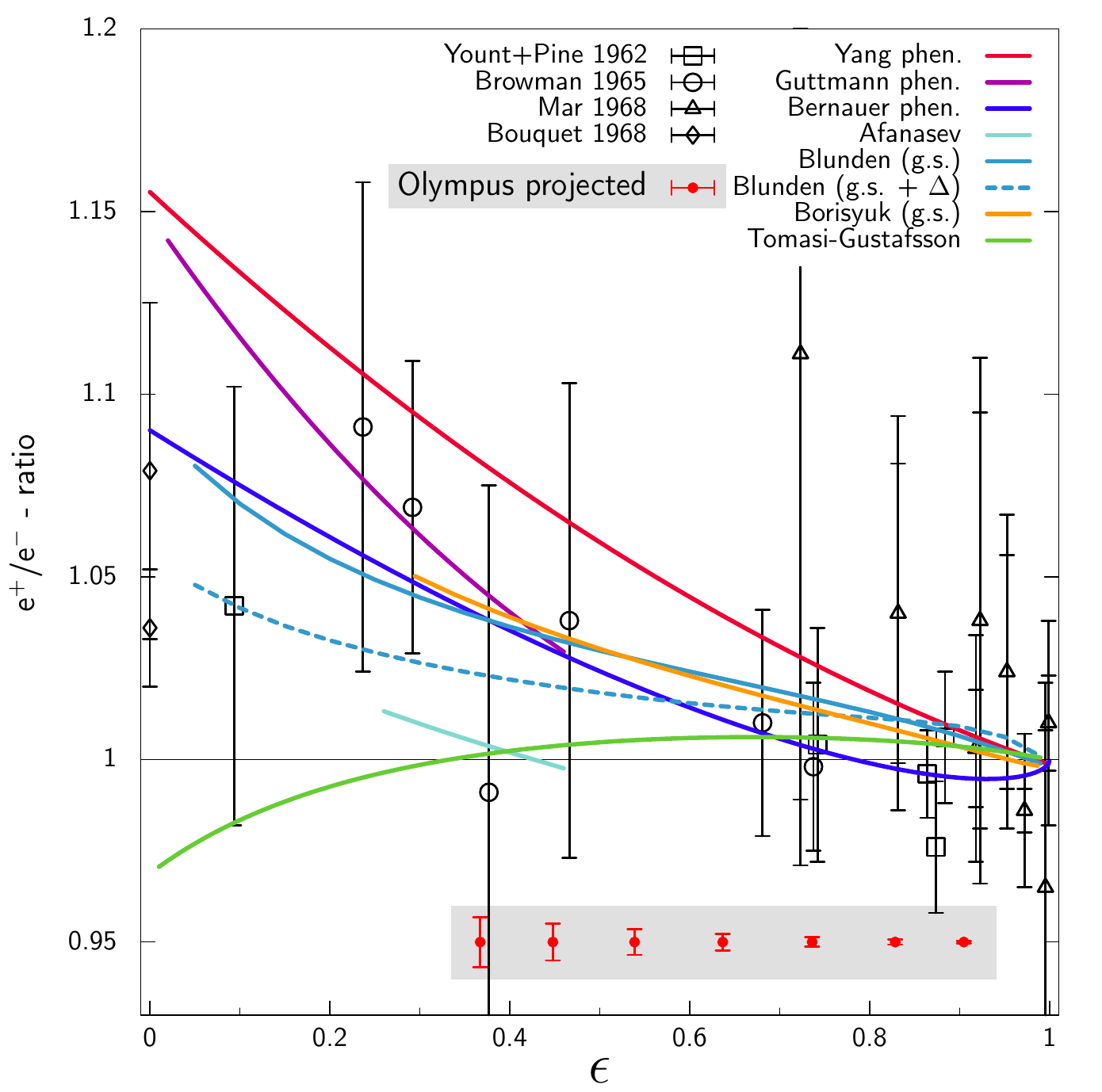}
\caption{The ratio of positron to electron elastic scattering cross
  sections at a beam energy of 2~GeV as a function of $\epsilon$
  showing phenomenological predictions~\citep{Bernauer:2013tpr,
    Chen:2007ac, Guttmann:2010au}, a selection of theoretical
  calculations of hard two-photon exchange~\citep{ Blunden:2003sp, 
    Chen:2004tw,Afanasev:2005mp,
    Blunden:2005ew, Kondratyuk:2005kk, Borisyuk:2006fh,
    TomasiGustafsson:2009pw}, and the projected OLYMPUS
  uncertainties. Also shown are existing ratio data~\citep{
    Yount:1962, Browman:1965zz, Mar:1968qd, Bouquet1968178} that were
  not taken at 2 GeV, plotted at their corresponding values of
  $\epsilon$.}
\label{fig:epratiow}
\end{figure}

To address this question, the OLYMPUS experiment was proposed to
measure the ratio between the positron-proton and electron-proton
elastic scattering cross sections. In the single-photon exchange
approximation this ratio is unity.  At next-to-leading order, the
interference of the one-photon and two-photon exchange diagrams
changes sign between electron and positron scattering. The two photon
exchange effect is expected to depend on the lepton scattering angle,
$\theta$, or virtual photon polarization, $\epsilon =
[1+2(1+\frac{Q^2}{4M_p^2})\tan^2(\theta/2)]^{-1}$, where $M_p$ is the
proton mass.  Measurements from the 1960s indicated some deviation in
the ratio from unity, but the uncertainties were large, as can be seen
in Fig.~\ref{fig:epratiow}.
OLYMPUS seeks to measure the ratio with uncertainty of less than 1\%
over the range 0.4~(GeV/c)$^2 \leq Q^2 \leq2.2$~(GeV/c)$^2$ for a single
beam energy $E = 2.01$~GeV..

OLYMPUS was approved for three months of dedicated operation at the
DORIS electron/positron storage ring at DESY, in Hamburg, Germany.
Beams of electrons or positrons were directed on an internal hydrogen
gas target, with the scattered leptons and recoiling protons detected
in coincidence over a wide range of scattering angles
($25\degree\leq\theta\leq75\degree$,
$-15\degree\leq\phi\leq15\degree$). The target was designed and built
at MIT and installed on the DORIS ring. The former BLAST detector was
shipped from MIT-Bates to DESY and placed around the target. The
detector used a toroidal magnetic field with a left/right symmetric
arrangement of tracking detectors and time of flight scintillators. In
addition, three new detector systems were designed and built to
monitor the luminosity during the experiment. Telescopes mounted at
$\theta = 12\degree$ consisted of triple GEM detectors from Hampton
University with readout electronics from INFN Rome and MWPC detectors
from PNPI. Symmetric M{\o}ller/Bhabha calorimeters from Mainz were
positioned at $1.29\degree$. The Bonn group provided the software and
hardware for the data acquisition system. The trigger and slow control
systems were developed by MIT.

The OLYMPUS experiment collected data in two periods: the February
period (January 20 - February 27, 2012) and the fall period (October
24, 2012 - January 2, 2013). During the February period, the beam
species was typically changed daily, and the magnet polarity was
changed randomly, but equally, every 6 hours. For the February data
run, there was a leak in the target gas supply that caused only a
fraction of the measured flow to reach the target cell. Because of
this, a lower than expected luminosity was obtained. The gas leak was
repaired in the summer so that it was possible to achieve high
luminosity in the fall period. However, it was discovered that at high
luminosity and negative magnet polarity too many electrons were bent
into the drift chambers, preventing their operation. After several
tests and attempts to remedy this, it was decided to operate at high
luminosity but primarily with positive magnet polarity for most of the
fall period.

The following sections describe the accelerator, target, detectors,
data acquisition, and operation in more detail.

\section{DORIS Storage Ring at DESY}
\label{sec:doris}
 
The DORIS storage ring at DESY originally began operation in 1974 as
an electron-electron and electron-positron collider. After its long
and successful operation for particle physics research, DORIS was
dedicated to synchrotron radiation studies in 1993. Since DORIS had
access to both positron and electron sources and could circulate both
species at several GeV energies, it met the requirements for the
OLYMPUS experiment. Additionally, the infrastructure at the location
in the beamline of the former ARGUS experiment~\citep{Albrecht:1996gr}
provided an excellent match to the size and needs of OLYMPUS. In 2009,
the shutdown of DORIS was scheduled for the end of 2012, placing a
tight time constraint on OLYMPUS.
 
Although the DORIS accelerator and the ARGUS detector site were well
suited to the OLYMPUS experiment, several modifications were
required. In particular, a number of considerations were necessary to
allow DORIS to continue to operate as a synchrotron light source after
OLYMPUS was installed (although not during OLYMPUS data taking).
These included:
\begin{itemize}
\item[-] RF cavities that had been installed at the detector site had
  to be relocated 26~m upstream.
\item[-] An additional quadrupole was installed on each side ($\pm 7$
  m) of the OLYMPUS interaction region to reduce the beam size for the
  OLYMPUS target while not significantly affecting the beam profile in
  synchrotron radiation source elements. This was necessary due to the
  impracticality of removing the OLYMPUS target for synchrotron
  runs.
\item[-] The OLYMPUS target required cooling during synchrotron
  radiation runs due to the wakefield heating caused by the 150 mA,
  4.5 GeV, 5-bunch beam.
\item[-] A number of tests and improvements were required to achieve
  the 10-bunch, 2.01 GeV beam conditions for OLYMPUS operation with
  adequate currents and lifetimes, including the implementation of a
  multi-bunch feedback system.
\end{itemize}

A key feature of the OLYMPUS experiment was the frequent switching
between electron and positron beams. The DORIS pre-accelerators were
already able to switch between electrons and positrons within
approximately 10 minutes, but the extraction from the pre-accelerators
to DORIS, the transport line, and the DORIS ring needed several
modifications:
\begin{itemize}
\item[-] The high voltage pulse power supplies for the pre-accelerator
  extraction and the DORIS injection kickers had to be rebuilt.
\item[-] The septa magnets for pre-accelerator extraction and DORIS
  injection were modified to serve as bipolar devices.
\item[-] Remotely-controlled polarity switches for a number of 800 A
  magnet power supplies had to be constructed and installed.
\end{itemize}

The daily switching of the beam species for OLYMPUS posed a challenge
during the fall period when DORIS and the PETRA storage ring operated
in parallel. The two rings shared the same pre-accelerators, and PETRA
only circulated positrons. The procedure for switching the polarity of
the pre-accelerators was optimized so that PETRA could be refilled
with positrons in approximately five minutes, causing only a small
delay in electron refills for DORIS.

Since the injection into DORIS occurred at full energy, it was
possible to run in top-up mode.  This allowed OLYMPUS to operate with
a higher target density while maintaining a high average beam current,
while also keeping the beam current at a more constant level.  The
injection process was optimized in order to minimize beam losses,
which prevented excessive rates and high voltage trips in the OLYMPUS
detectors.

The radiation levels in the region downstream of the experiment
increased when gas was added to the target, and additional shielding
was installed to account for this. Also, the beam scrapers upstream of
the experiment were optimized to minimize the noise rates in the
experiment.

To monitor the beam energy, a dipole reference magnet was installed in
series with the DORIS dipole magnets.  This magnet included a rotating
coil to measure the integrated field strength. The accelerator archive
system monitored all relevant data, power supply currents for all
magnets, beam position data, scraper positions, etc.\@ and provided
much of this information to the OLYMPUS slow control system.

\section{Target and Vacuum Systems}
\label{sec:target}
 
The OLYMPUS experiment used an unpolarized, internal hydrogen gas
target cooled to below 70~K. The hydrogen gas flowed into an
open-ended, 600~mm long, elliptical target cell (Sec.~\ref{sec:cell}). The
target cell was housed in a scattering chamber (Sec.~\ref{sec:chamber})
that had thin windows between the cell and the detectors. The target
system was designed to withstand both OLYMPUS running conditions and
those when DORIS operated as a synchrotron source.  A series of
wakefield suppressors (Sec.~\ref{sec:wakefield}) were necessary to reduce
the heat load on the target cell.  A tungsten collimator
(Sec.~\ref{sec:collimator}) was also housed in the scattering chamber to
prevent synchrotron radiation, beam halo, and off-momentum particles
from striking the target cell.  Finally, an extensive vacuum system
(Sec.~\ref{sec:vacuum}) of turbomolecular and Non-Evaporable Getter (NEG)
pumps was employed to preserve the vacuum in the DORIS storage ring.

\subsection{Target Cell}
\label{sec:cell}

The target cell consisted of an open-ended, elliptical cylinder (27~mm
horizontal~$\times$~9~mm vertical~$\times$~600~mm long) made from
0.075~mm thick aluminum.  The elliptical shape was chosen to match the
DORIS beam envelope and was set to approximately the $10\sigma$
nominal horizontal and vertical beam width at the OLYMPUS interaction
point to minimize the amount of beam halo striking the cell walls.

\begin{figure}[htbp]
\centering 
\includegraphics[width=\columnwidth]{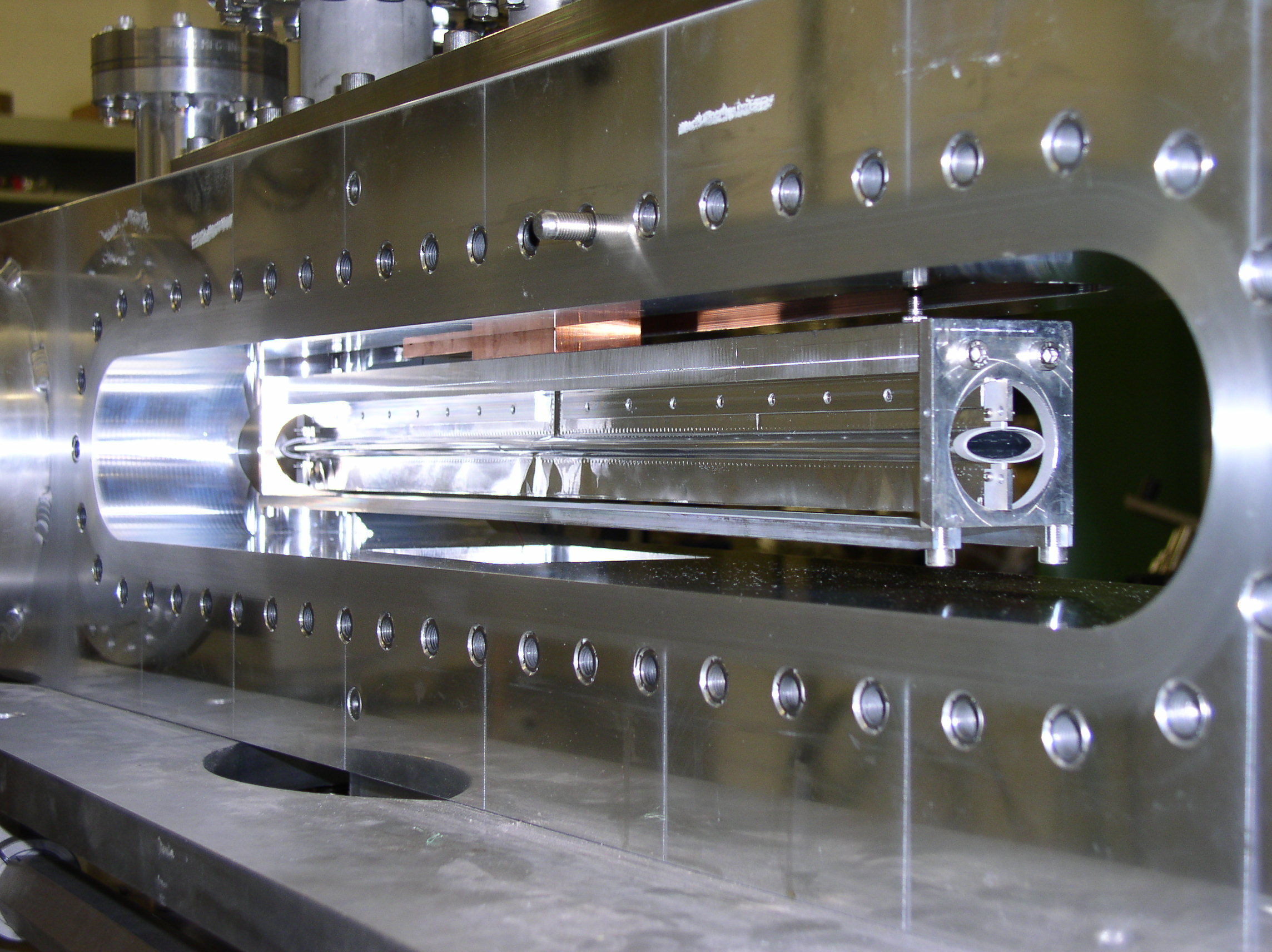}
\caption{Photograph of one of the OLYMPUS target cells mounted inside
  the scattering chamber.}
\label{fig:targetcell}
\end{figure}

The INFN Ferrara group produced several target cells for the OLYMPUS
experiment.  Cells were formed from two identical stamped sheets of
aluminum that were spot-welded together along the top and bottom
seams. Each cell was mounted in a frame by a clamp that ran the entire
length of the top seam. The frame was made of 6063 aluminum to provide
high thermal conductivity at cryogenic temperatures. When installed in
the scattering chamber, the cell and frame assembly was suspended from
a flange in the top of the scattering chamber (shown in
Fig.~\ref{fig:targetcell})
and its position and orientation could be adjusted. The entire cell
and frame assembly was cooled by a cryogenic coldhead. The assembly
was wrapped in several layers of aluminized mylar to insulate it from
thermal radiation. Without beam or gas flow, the target could reach
temperatures below 40~K. During high-luminosity running, a temperature
of about 70~K was sustained.

During operation, hydrogen gas was flowed through the target cell. The
hydrogen gas was produced by a commercial hydrogen generator and was
controlled by a series of valves, buffer volumes, and mass flow
controllers.  The gas entered the cell at the center, from a tube that
fit snuggly into an opening of the cell's top seam. The gas diffused
outwards to the open ends of the cell, where it was removed by the
vacuum system. This diffusion was slowed because the hydrogen quickly
cooled to the temperature of the cell. The density distribution in the
cell was triangular, with peak density at the center of the cell
falling to zero density at either end. A flow rate of $1.5 \times
{10}^{17}$ H$_2$ molecules per second was required to produce a target
thickness of $3 \times {10}^{15}$~atoms~cm$^{-2}$.

\subsection{Scattering Chamber}
\label{sec:chamber}

\begin{figure}[htbp]
\centering 
\includegraphics[width=\columnwidth]{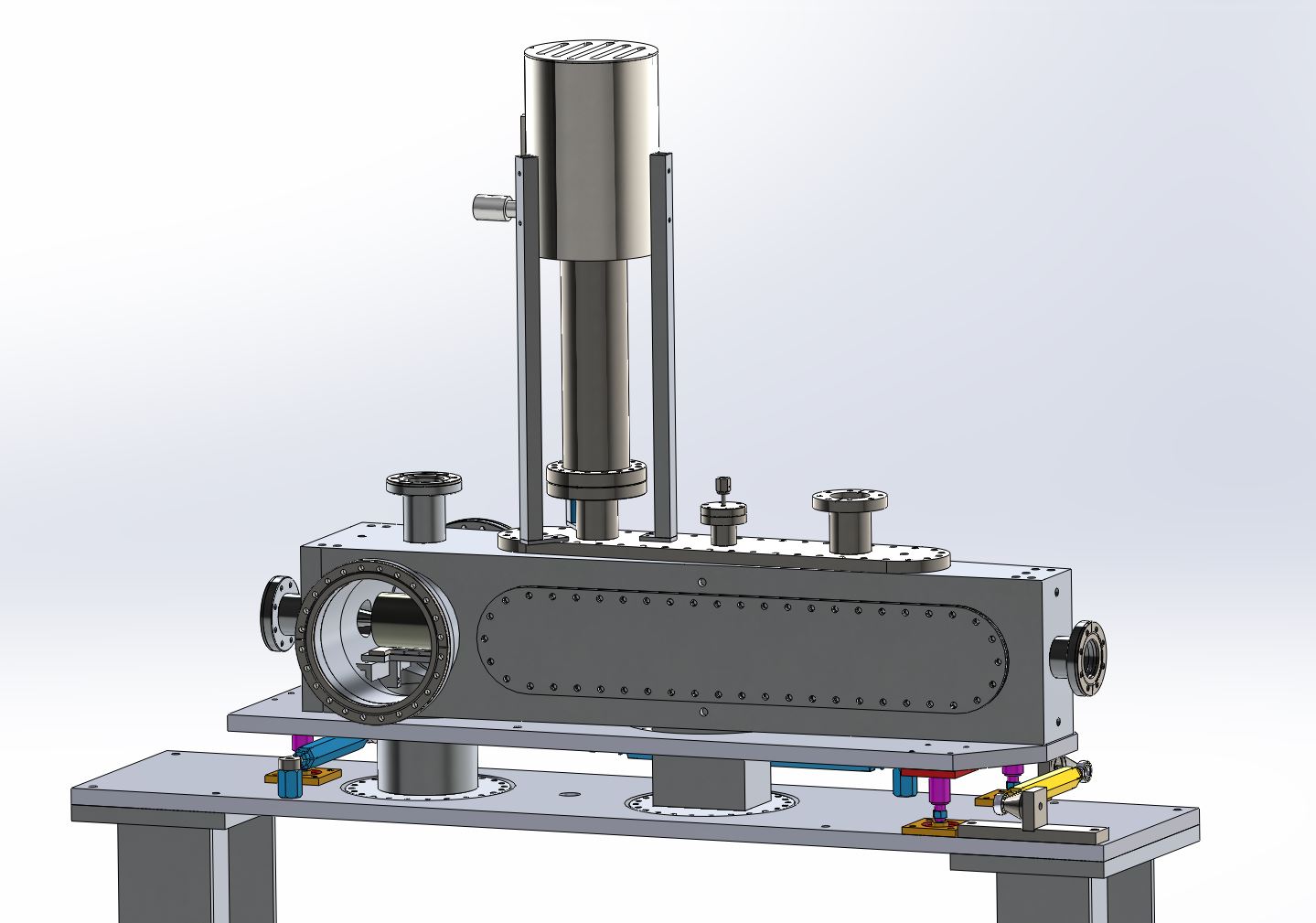}
\caption{CAD model of the OLYMPUS scattering chamber.}
\label{fig:chamber}
\end{figure}

The OLYMPUS scattering chamber (shown in Fig.~\ref{fig:chamber})
was 1.2~m long and was machined from a solid block of aluminum, with
large area windows on the left and right faces. The windows were made
of 0.25~mm thick 1100 aluminum, and nominally subtended a polar
angular range of $8\degree$ to $100\degree$ from the center of the
target, $6\degree$ to $90\degree$ from 200~mm upstream, and
$10\degree$ to $120\degree$ from 200~mm downstream.  The chamber was
trapezoidal in shape to make more of the target cell ``visible'' to
the $12\degree$ detectors.

\begin{figure}[htbp]
\centering 
\includegraphics[width=\columnwidth]{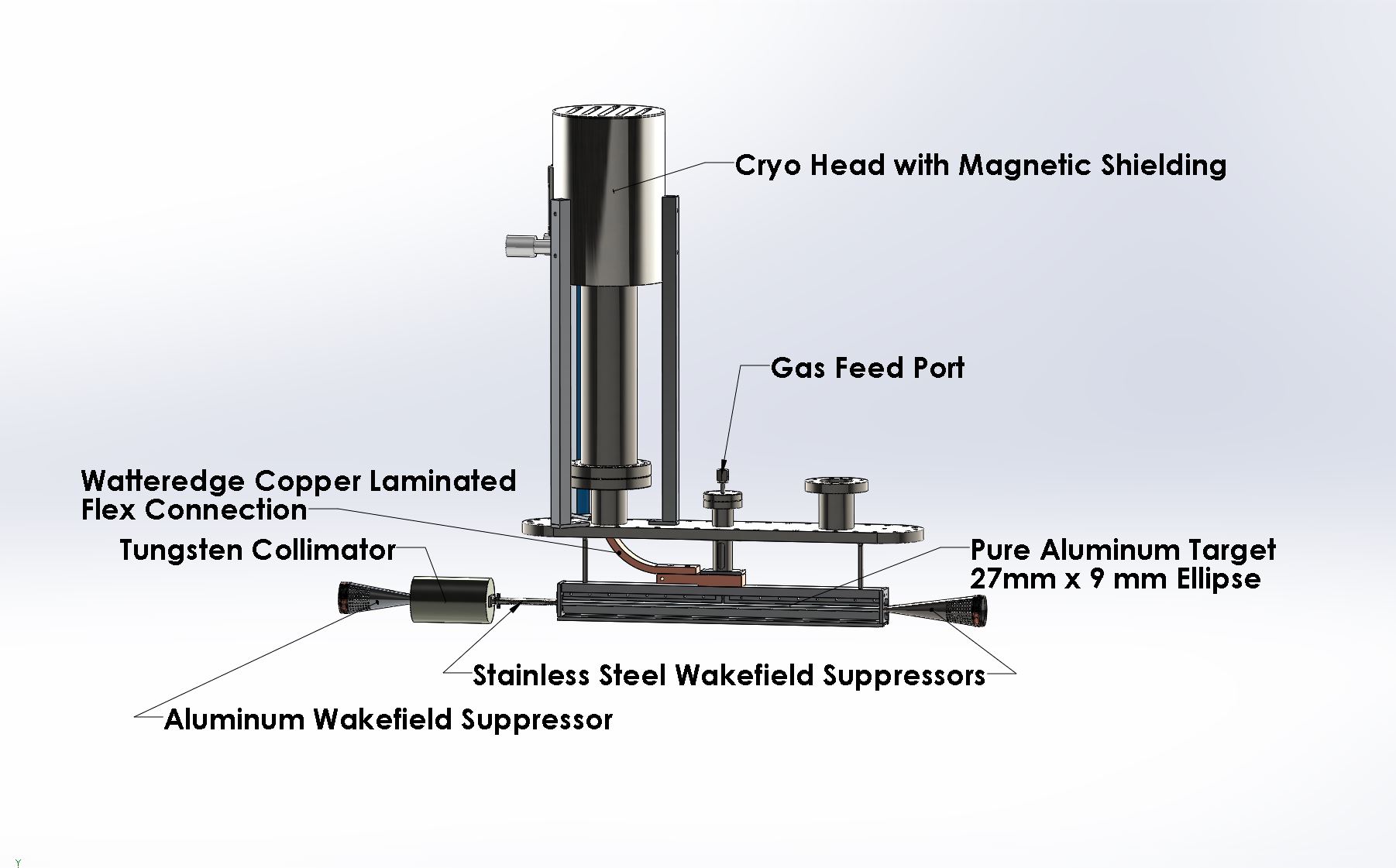}
\caption{CAD model of the target cell, wakefield suppressors, and
  collimator inside the OLYMPUS scattering chamber.}
\label{fig:target}
\end{figure}

In addition to windows, the chamber had ports for the beamline (up-
and downstream), for pumping (on the bottom surface), for access to
the collimator (on the left and right), and for the target cell flange
on the top, which had feedthroughs for the hydrogen gas, the coldhead,
and various sensors.  The main components inside the scattering
chamber are shown in Fig.~\ref{fig:target}.

\subsection{Wakefield Suppressors}
\label{sec:wakefield}

Wakefield suppressors were necessary to maintain the target cell at
cryogenic temperatures by preventing heating caused by wakefields. The
wakefield suppressors consisted of conducting transitions that were
added to fill gaps between conducting structures surrounding the beam.
Any sharp transitions or gaps in conductivity would act as electrical
cavities that would be excited by the passing beam, creating
wakefields and producing heat. To prevent this, three wakefield
suppressors were produced to cover the following three transitions:
\begin{enumerate}
\item from the circular upstream scattering chamber port (60~mm in
  diameter) to the 25~mm $\times$ 7~mm elliptical opening of the
  collimator,
\item from the exit of the collimator to the entrance of the target
  cell (both 27~mm $\times$ 9~mm ellipses), and
\item from the 27~mm $\times$ 9~mm elliptical exit of the target cell
  to the circular downstream scattering chamber port (60~mm in
  diameter).
\end{enumerate}
With these wakefield suppressors, a target temperature of around 50~K
was maintained during synchrotron operation, and a temperature less
than 70~K was maintained during high-luminosity OLYMPUS running.

\begin{figure}[htbp]
\centering 
\includegraphics[width=\columnwidth]{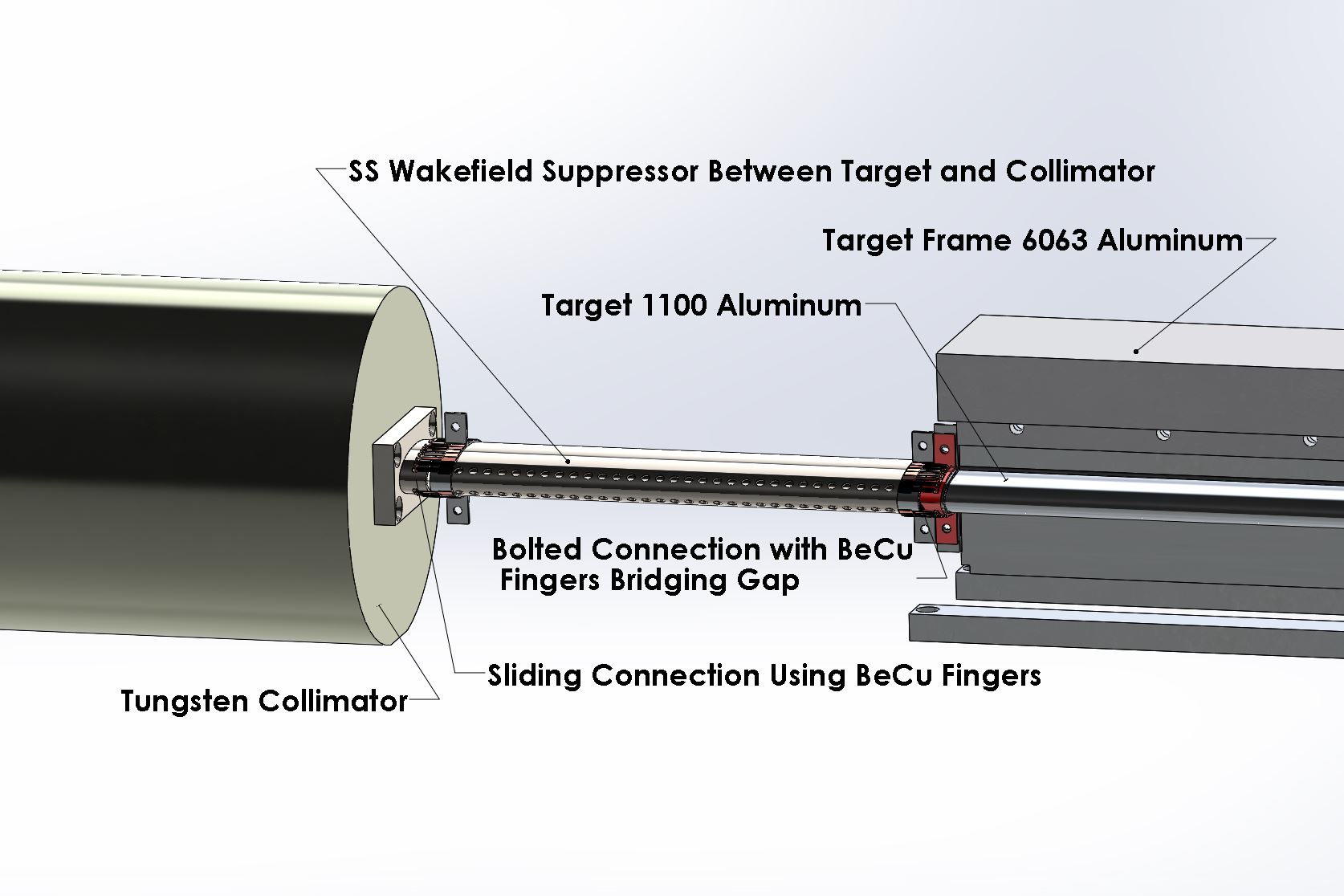}
\caption{CAD model of the wakefield suppressor between the collimator
  and the target cell.}
\label{fig:wakefield}
\end{figure}

The wakefield suppressors were made of stainless steel (except the
upstream wakefield suppressor, which was made of aluminum) and plated
with silver for improved electrical conductivity. The surfaces were
smooth except for many small holes, which were drilled to allow the
vacuum system to pump gas through them. The ends of the wakefield
suppressors had beryllium-copper spring fingers around their
circumference. These spring fingers made sliding connections at an
interface that allowed for thermal expansion while maintaining good
electrical contact. The upstream wakefield suppressor was screwed
directly to the collimator making a sliding connection with the
upstream scattering chamber port. The other two wakefield suppressors
were fixed to rings clamped to the ends of the target making sliding
connections to either the downstream scattering chamber port or the
collimator.  A close-up view of the middle wakefield suppressor is
shown in Fig.~\ref{fig:wakefield}.

\subsection{Collimator}
\label{sec:collimator}

Fig.~\ref{fig:wakefield} also shows the fixed collimator in front of the
target cell.  The collimator consisted of a 139.7~mm long cylinder of
tungsten 82.55~mm in diameter. The outer dimensions were chosen after
performing a study on simulated showers of beam-halo particles. It had
a tapered elliptical aperture with entrance 25~mm $\times$ 7~mm and
exit 27~mm $\times$ 9~mm.  The collimator was machined from a solid
block of tungsten using wire electrical discharge
machining\footnote{Jack's Machine Co. Hanson, MA 02341}.  The entrance
dimensions were chosen to be slightly smaller than those of the
storage cell to shield the target cell walls.

\subsection{Vacuum System}
\label{sec:vacuum}

A system of six magnetic levitation turbomolecular pumps
(Osaka\footnote{Osaka Vacuum Ltd., Osaka, Japan} TG 1100M and
Edwards\footnote{Edwards, Crawley, UK} STP 1003C, 800 L/s capacity)
and NEG pumps (SAES\footnote{SAES Group, Lainate, Italy} CapaciTorr
CFF 4H0402, 400~L/s capacity) was used to pump the section of beamline
inside the OLYMPUS experiment.  This system utilized three stages of
pumping to reduce the pressure from the relatively high pressure
($\sim10^{-6}$~Torr) at the scattering chamber (caused by hydrogen gas
flowing into the target cell) to the low pressure ($\sim10^{-9}$~Torr)
of the DORIS storage ring.

\begin{figure}[htbp]
\centering 
\includegraphics[width=\columnwidth]{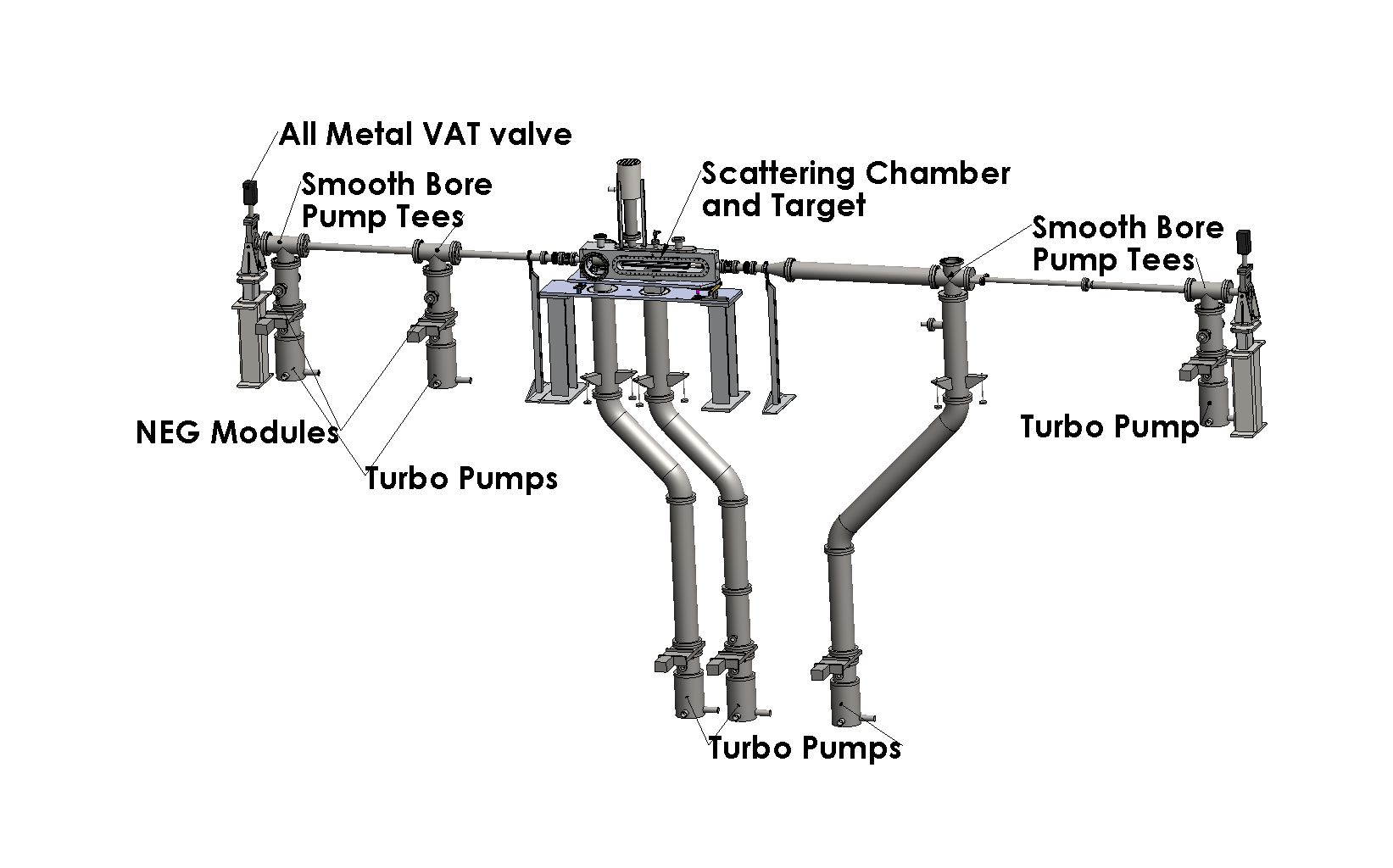}
\caption{CAD model of the vacuum system employed for the OLYMPUS
  experiment.}
\label{fig:vacuum}
\end{figure}

The vacuum system is shown in Fig.~\ref{fig:vacuum}.
Two turbo pumps located in the pit beneath the experiment were
directly connected to the scattering chamber through 200~mm diameter
pipes.  Two more turbo pumps were connected to the up- and downstream
beamlines approximately 2~m from the target. At approximately 3~m from
the target another two turbo pumps were used to reduce the pressure in
the beamline to the level acceptable for the DORIS storage ring.  The
four pumping stations furthest from the target also had NEG pumps to
improve the pumping of hydrogen.

\section{The OLYMPUS Detector}
\label{sec:detector}

\begin{figure}[htbp]
\centering
\includegraphics[width=\columnwidth]{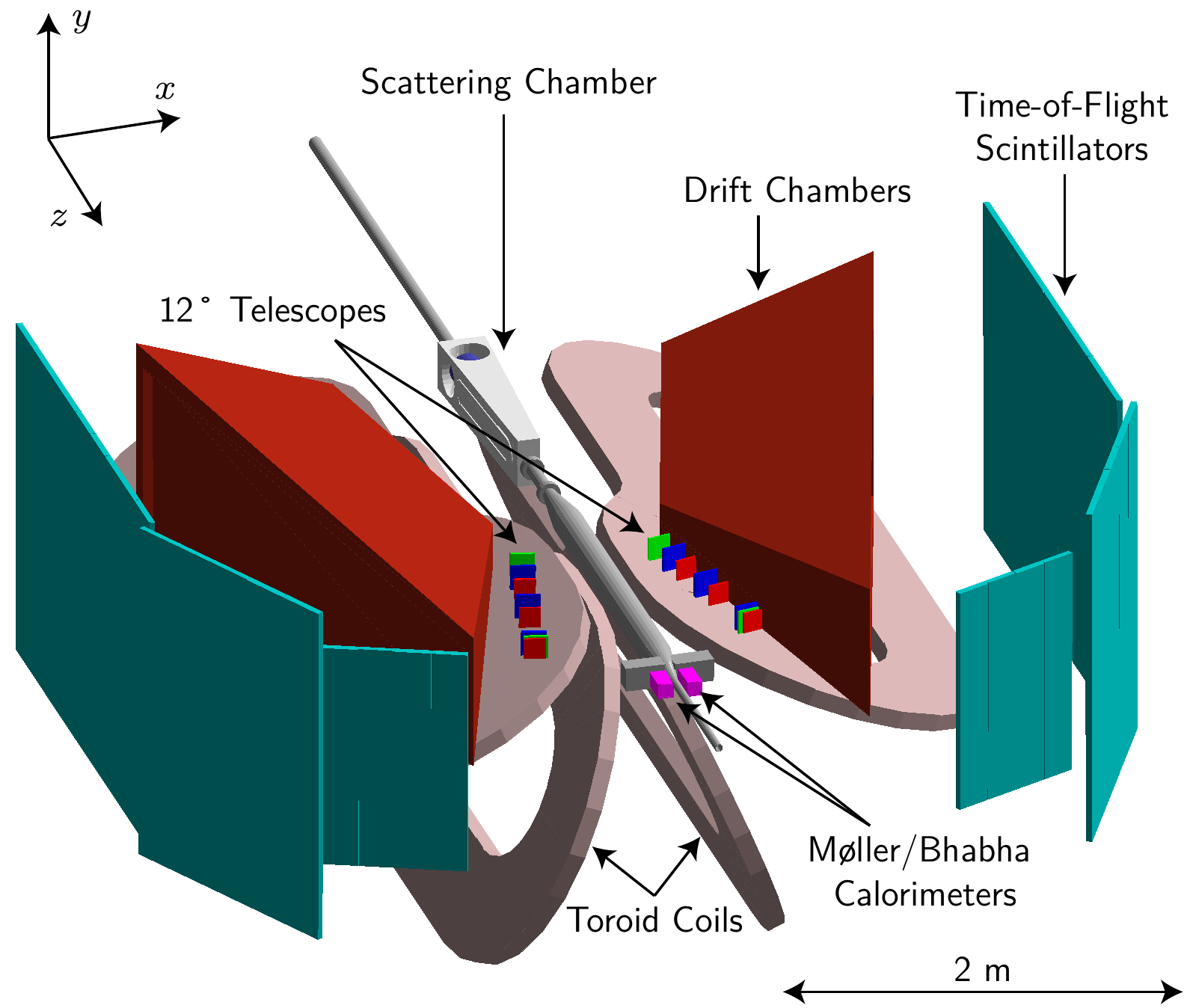}
\caption{A solid-model representation of the OLYMPUS detector with the
  top four magnet coils removed to show the instrumented horizontal
  sectors.}
\label{fig:det_overview}
\end{figure}

The OLYMPUS spectrometer consisted of an eight-coil toroidal magnet
with detectors in the two horizontal sectors on either side of the
beamline (see Fig.~\ref{fig:det_overview}).
Each of these sectors contained drift chambers for particle tracking
and an array of time-of-flight scintillator bars for triggering and
measurements of energy deposition, particle position, and timing. To
monitor the luminosity, OLYMPUS had a redundant system consisting of
symmetric M{\o}ller/Bhabha (SYMB) calorimeters at $\theta =
1.29\degree$ and detector telescopes at $12\degree$ in both sectors,
each consisting of three gas electron multiplier (GEM) detectors
interleaved with three multi-wire proportional chambers (MWPCs).
 
The toroidal magnet, drift chambers, time-of-flight detectors, support
frames, and many of the readout and control electronics were
originally part of the BLAST spectrometer~\citep{Hasell:2009zza} at
MIT-Bates.  These components were shipped to DESY in spring 2010 where
they were reassembled, reconditioned, and modified as necessary for
installation in the OLYMPUS detector.

The OLYMPUS experiment was installed in the straight section of the
DORIS storage ring, in the location of the former ARGUS
experiment~\citep{Albrecht:1996gr}. The initial assembly took place
from June 2010 to July 2011 outside of the DORIS tunnel, to avoid
interferring with DORIS operation. The detector was assembled on a set
of rails that led (through a removable shielding wall) to the ARGUS
site. When the assembly was complete, the shielding wall was removed,
the spectrometer was rolled into place in the tunnel, and the wall was
rebuilt. The experimental site was 7~m wide, with a 5~m deep pit below
the beam height. The pit was a convenient location for vacuum pumps,
power supplies, and the target gas system because it was deep enough
to be outside of the fringes of the magnet field.

In the area outside the shielding wall was an electronics hut, which
was supported on the same set of rails. The hut housed the detectors'
readout and control electronics, the high voltage supplies, and the
computer systems. The electronics hut could be accessed even when the
DORIS beam was circulating.

The following sections describe the detector components in greater
detail.

\subsection{Toroidal Magnet}
\label{sec:magnet}

\begin{figure}[htb] 
\centering
\includegraphics[width=\columnwidth]{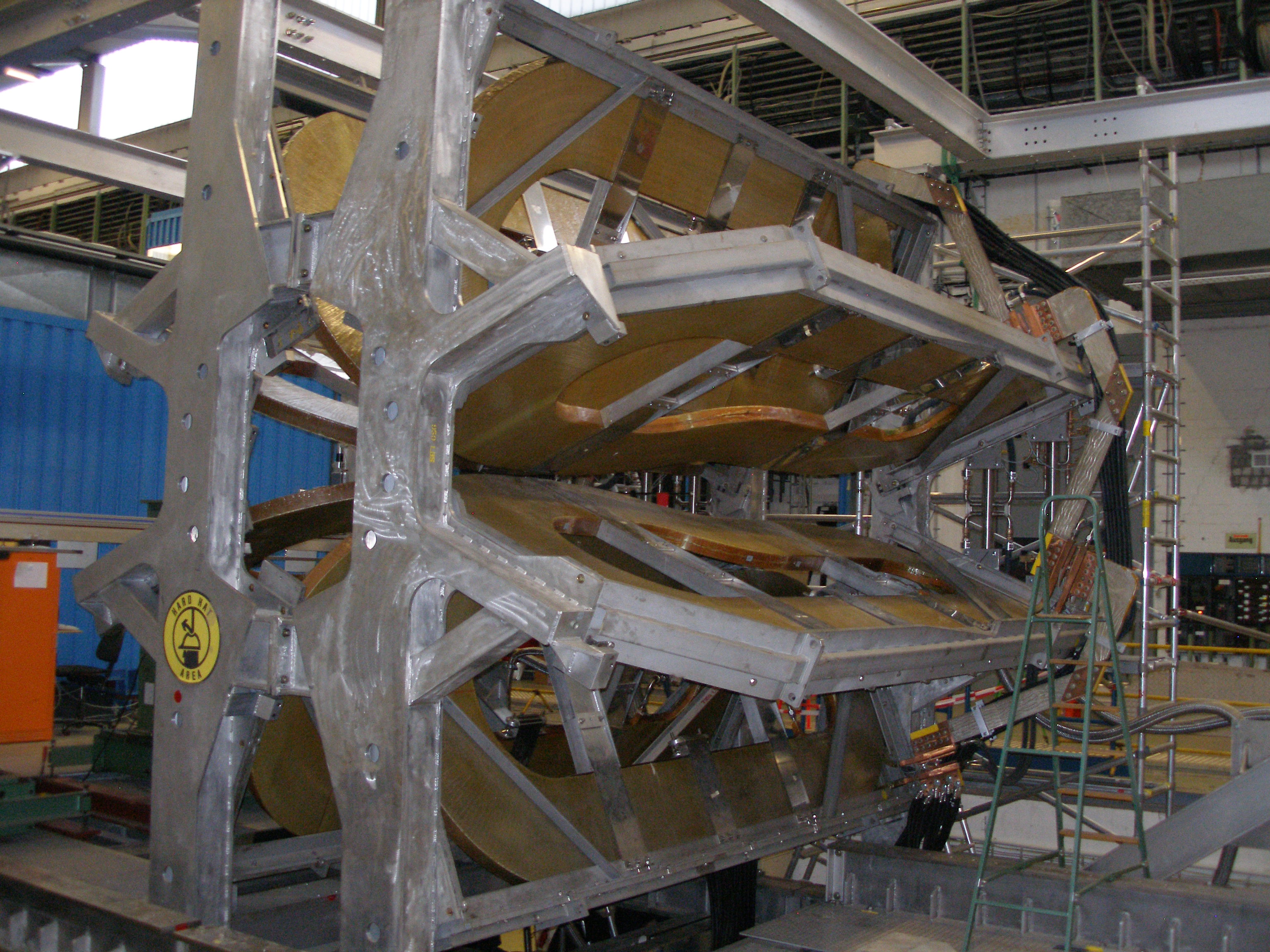}
\caption{The toroid magnet assembled at DESY before the subdetectors
  were installed}
\label{fig:toroid}
\end{figure}

The toroidal magnet consisted of eight copper coils placed around the
beam line and scattering chamber so that the beam traveled down the
toroid's symmetry axis (see Fig.~\ref{fig:toroid}).
The coils divided the space around the beamline into eight
sectors. The two sectors in the horizontal plane were instrumented
with detectors. During normal operation, the magnet produced a field
of about 0.28~T in the region of the tracking detectors.

The magnet was originally designed and used for the BLAST experiment,
and has been described in a previous article~\citep{Dow:2009zz}.  The
choice of a toroidal configuration for BLAST was made to ensure a
small field along the beamline in order to minimize any effects on a
spin-polarized beam and to limit field gradients in the region of the
polarized target.  Since OLYMPUS used neither a polarized beam nor a
polarized target, these concerns were not as important. However,
during the initial set-up, the magnetic field along the beamline was
measured and the coil positions adjusted to achieve an integrated
field $<0.005$~T$\cdot$m to avoid perturbing the beam's position or
direction.

\begin{figure}[htb] 
\centering 
\includegraphics[width=\columnwidth]{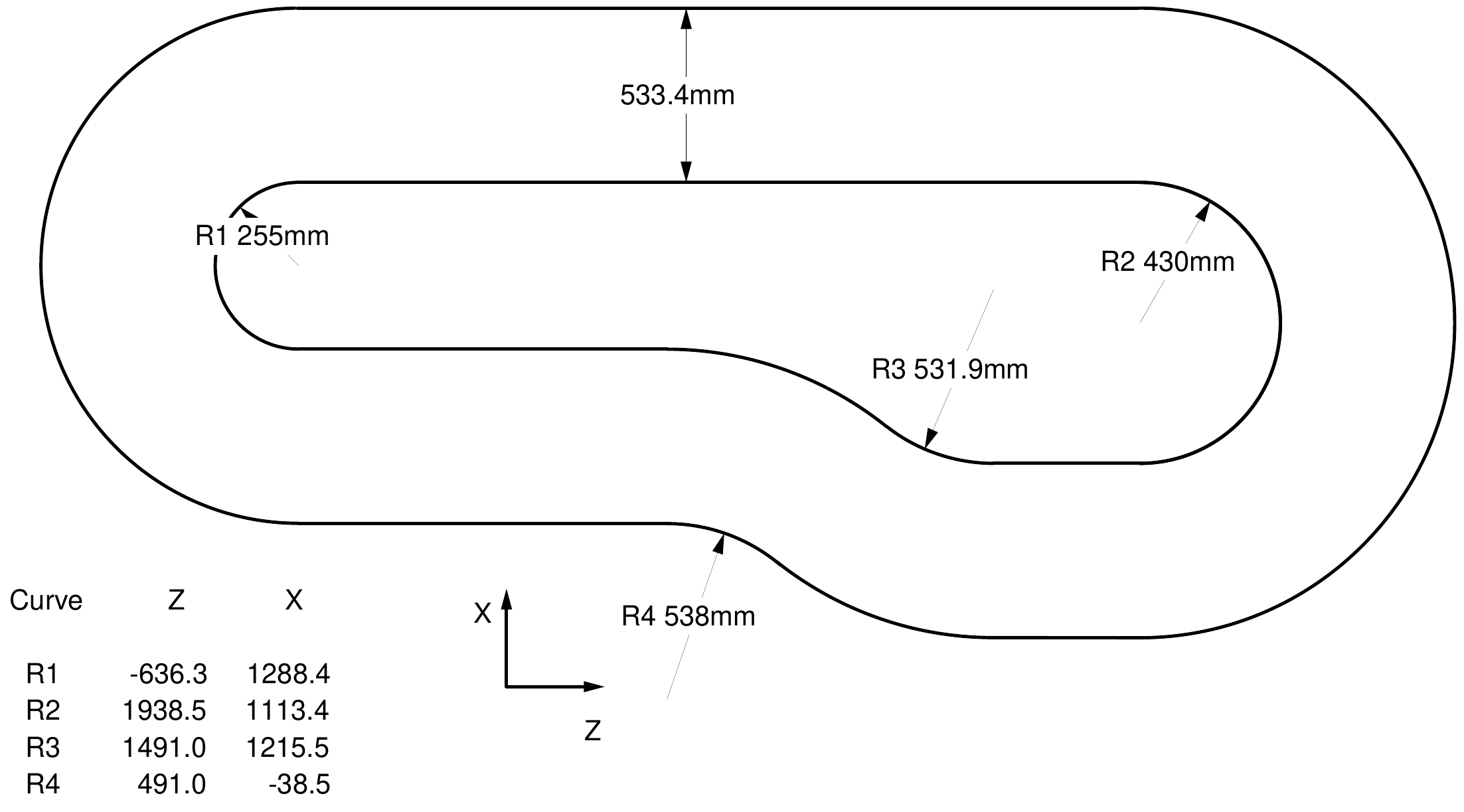}
\caption{Planar view of BLAST coil outline showing dimensions and
  position relative to the center of the target cell.}
\label{fig:coil}
\end{figure}

Each of the toroid's eight coils consisted of 26~turns of 1.5~inch
square copper tubes, organized into two layers of 13~turns. A circular
hole, 0.8~inches in diameter, ran down the length of each tube and
served as a conduit for cooling water. During assembly, the tubes were
individually wrapped with fiberglass tape and then collectively potted
in an epoxy resin matrix. The final outline and nominal position
relative to the beam line and target center at the coordinate origin
are shown in Fig.~\ref{fig:coil}.
 The coils are narrower at one end to
accommodate the scattering chamber and wider at the other to extend
the high-field region to more forward angles, where scattered
particles have higher momenta.

The magnetic field served two purposes. The first was to bend the
tracks of charged particles, allowing their momentum and charge sign
to be determined from the curvature of their tracks. The second was to
sweep away low-energy, charged background particles from the tracking
detectors. Though a stronger magnetic field would have improved
momentum resolution and reduced the background, it would also have
increased the Lorentz angle of drift electrons in the tracking
detectors, making track reconstruction more difficult. A balance was
struck by choosing a current of 5000~A for normal operation, which
produced a field of about 0.28~T in the high-field regions.

Originally, it was planned to alternate the polarity of the magnet
every few hours to reduce systematic uncertainties. However, this
proved impractical at high luminosity. In the negative polarity
setting, the magnet bent negatively charged particles outward from the
beamline. The drift chambers were hit with a large background of
low-energy electrons, which frequently caused the high voltage supply
to exceed its current threshold and trip. Attempts to adequately
shield the drift chambers, both by adding material and by increasing
the magnetic field strength, were unsuccessful. Consequently, the
negative polarity setting was limited to low-luminosity running, and
only about 13\% of the total luminosity was collected in this mode.
The limited negative polarity data will provide a check on systematic
uncertainties.

After the experimental running period was completed, the detectors and
downstream beamline were removed in order to conduct a measurement of
the magnetic field.  By convention, the direction of the beam was
labeled as the OLYMPUS $z$-axis, the $y$-axis pointed up, and the
$x$-axis pointed toward the left sector, forming a right-handed
coordinate system.  The field region was scanned using a 3D Hall probe
mounted to a rod, driven by several translation tables. The rod was
mounted to a long XYZ table with a range of motion of 0.2~m $\times$
0.2~m $\times$ 6~m.  This long table was supported by two large XY
tables that augmented the $x$ and $y$ ranges each by 1~m. The range of
motion was further extended in $x$ by substituting rods of different
lengths and in $y$ by adding a vertical extension piece. The apparatus
was used to measure the field over a grid of points on the left
sector, before being transported and reassembled for a similar
measurement of points on the right sector. The grid extended from
-0.5~m to 3.5~m in $z$. In $x$ and $y$, the grid was limited to the
triangular space between the coils, but extended to $\pm2.7$~m on
either side of the beamline. The grid points were spaced 0.05~m apart
in the region within 1~m of the beamline, and 0.10~m apart in the
outer region, where the field changed less rapidly. In total,
approximately 35,000 positions were measured over a two month period,
including the downstream beamline region, which was measured
redundantly from the left and the right.

\begin{figure}[htbp] 
\centering 
\includegraphics[width=\columnwidth]{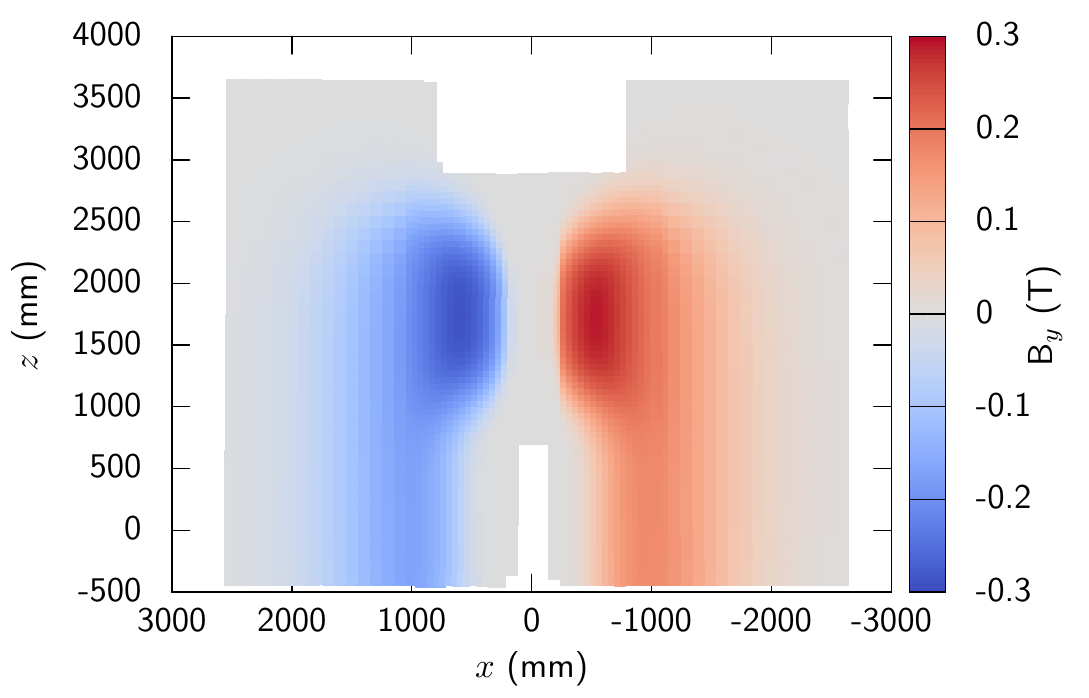}
\caption{Measurements of the vertical magnetic field component $B_y$
  in the horizontal plane as viewed from above}
\label{fig:field_data}
\end{figure}

After the initial setup of the apparatus, the precise position of the
XYZ tables was measured with a laser tracking station over the course
of a typical scan in $z$. This showed that the Hall probe position
varied in $x$ and $y$ as a function of $z$ during a scan, but that the
shape was quite reproducible. To correct for this variation, the start
and end points of each scan were measured using a theodolite and a
total station. This data then allowed the position of the Hall probe
to be determined for each measurement. Position-corrected data for the
vertical component of the field are shown in Fig.~\ref{fig:field_data}.

After correcting the Hall probe positions, a fit was performed to the
magnetic field data.  The fit was based on a model of the coil
geometry with a Biot-Savart calculation of the magnetic field. The fit
allowed the coil positions to vary slightly to best match the
measurements.  This model was then used to extrapolate the field over
the entire volume around the OLYMPUS detector for use in track
reconstruction and in the OLYMPUS Monte Carlo simulation.

\subsection{Drift Chambers}
\label{sec:wc}

The drift chambers used for the OLYMPUS experiment came from the BLAST
experiment at MIT-Bates and have been described in great detail
elsewhere~\citep{Hasell:2009zza}, so the following description will be
brief while mentioning new and updated features.

\begin{figure}[htbp]
\centering
\includegraphics[width=\columnwidth]{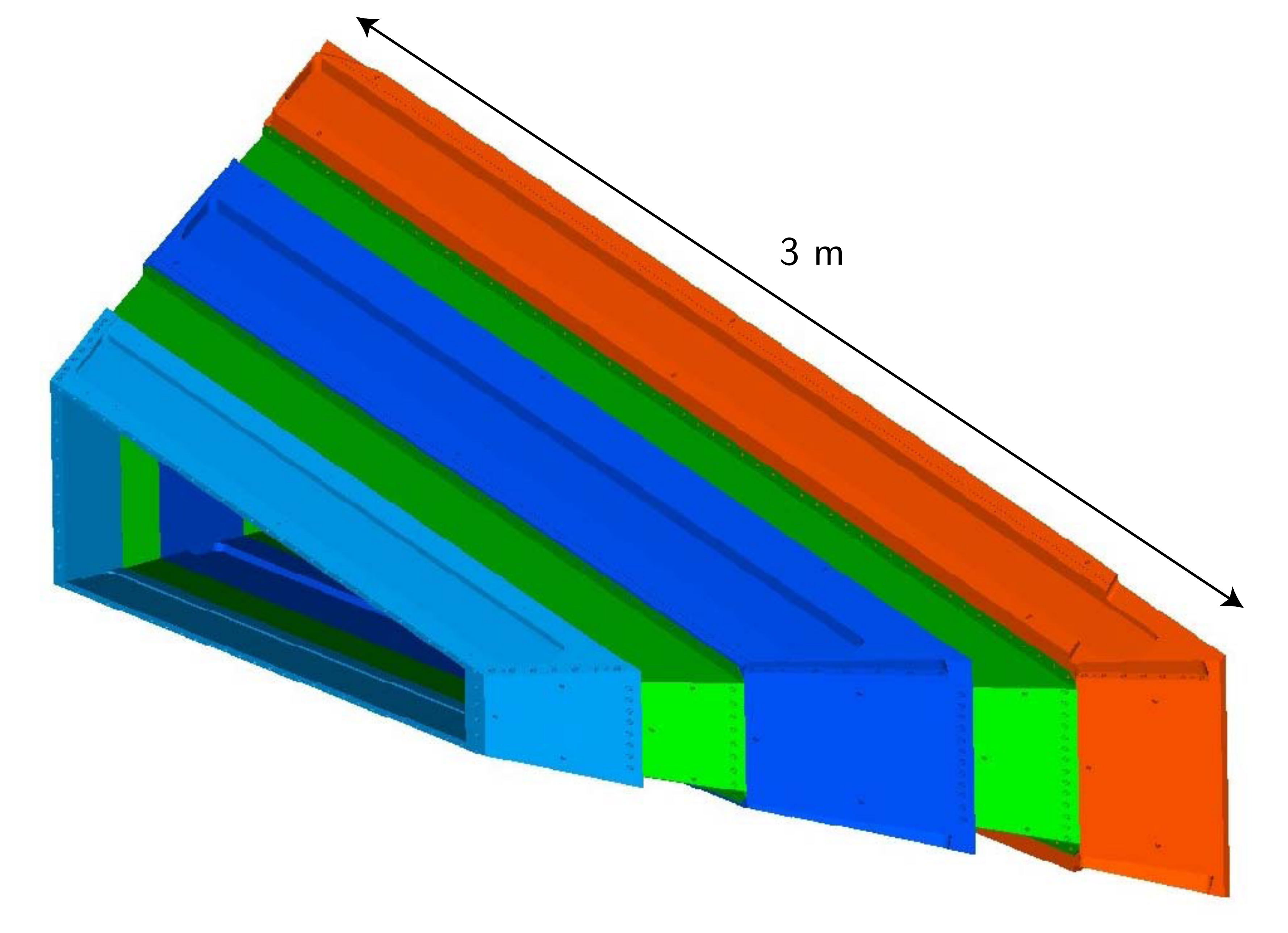}
\caption{Isometric view of all three drift chambers assembled into a
  single gas volume.}
\label{fig:assembly}
\end{figure}

The drift chambers were used to measure the momenta, charges,
scattering angles, and vertices of out-going charged particles. The
drift chambers had a large angular acceptance, subtending a range of
$20\degree$--$80\degree$ in polar angle and $\pm15\degree$ in
azimuth. The chambers were oriented to be normal to a polar angle of
$73.54\degree$. Because of these choices, the chambers were
trapezoidal in shape (see Fig.~\ref{fig:assembly}).

The drift chambers were arranged in two sectors that were positioned
on either side of the target, in the horizontal plane. Each sector
contained three drift chambers (inner, middle, and outer) joined
together by two interconnecting sections to form a single gas
volume. Thus, only one entrance and one exit window were needed,
reducing multiple scattering and energy loss. The drift chambers
combined had approximately 10,000 wires, which were used to create the
drift field.  Of these, 954 were sense wires, which read out the
signals from ionization caused by a charged particle track.

\begin{figure}[htbp]
\centering 
\includegraphics[width=\columnwidth,viewport=0 270 560 540,clip]
{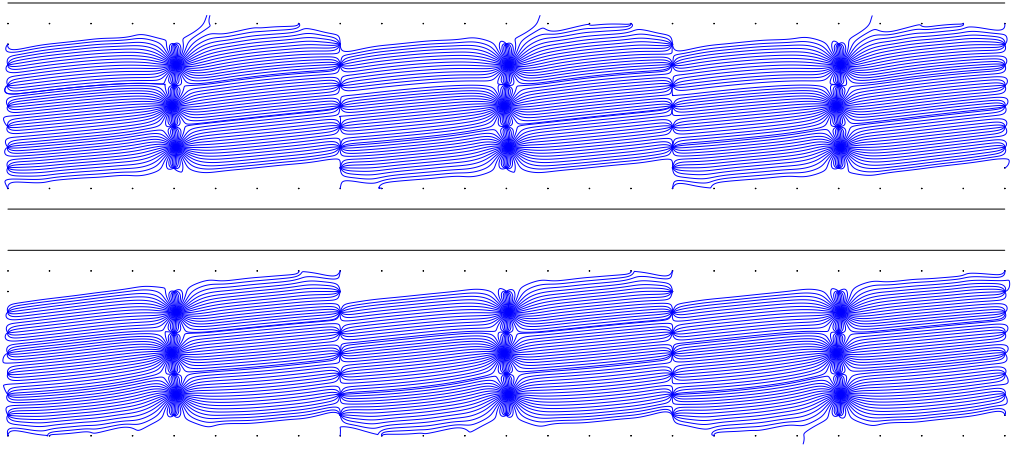}
\caption{Portion of a chamber showing the two super-layers of drift
  cells formed by wires. Lines of electron drift in the drift cells
  assuming a typical magnetic field around 3.0~kG are also shown.}
\label{fig:field}
\end{figure}

Each chamber consisted of two super-layers (or rows) of drift cells,
with 20~mm separation between the super-layers. The drift cells were
formed by wires in a ``jet style'' configuration.  Fig.~\ref{fig:field}
shows a cross-sectional view of a portion of one chamber with the two
super-layers of drift cells. It also shows characteristic
``jet-style'' lines of electron drift in a magnetic field.  Each drift
cell was $78\times40$~mm$^2$ and had three sense wires staggered
$\pm0.5$~mm from the center line of each cell to help resolve the
left/right ambiguity in determining position from the drift time.  The
wires in one super-layer were strung with a $10\degree$ stereo angle
relative to wires of the other so that each chamber could localize a
trajectory in three dimensions.

Because transporting the chambers in a way that would protect the
wires from breaking was infeasible, the chambers were completely
rewired in a clean room at DESY over a period of about three months
during the summer of 2010. In addition to new wires, improvements were
made to the front-end electronics, building on experience gained from
BLAST.

For the experiment, an Ar:CO$_2$:C$_2$H$_6$O gas mixture
(87.4:9.7:2.9) was chosen for the drift chambers.  The ethanol was
added by bubbling an Ar:CO$_2$ (90:10) gas mixture through a volume of
liquid ethanol kept near 5~C. The chambers were maintained at a
pressure of approximately 1~inch of water above atmospheric pressure
with a flow rate of around 5~L/min.

\begin{figure}[htbp]
\centering 
\includegraphics[width=\columnwidth]{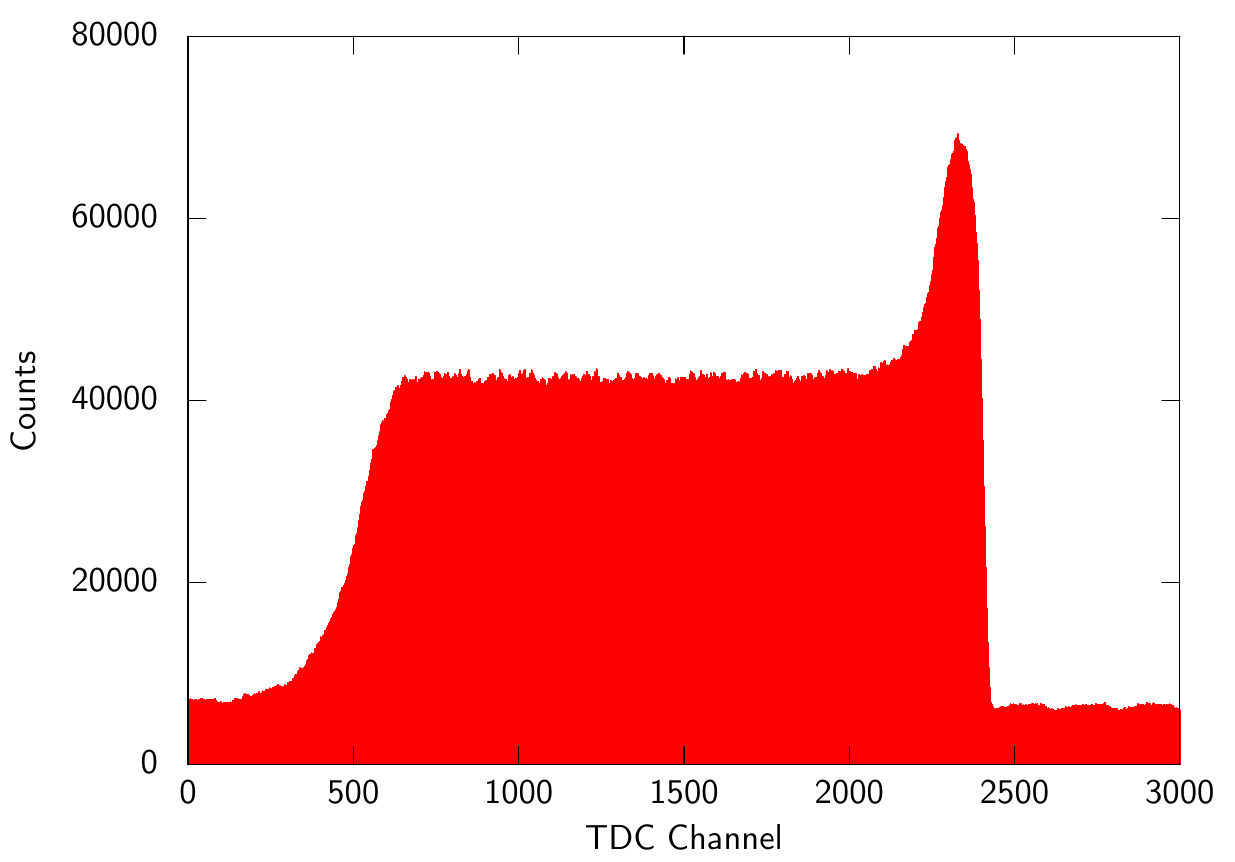}
\caption{A typical TDC spectrum for a single wire has a ``church
  shape,'' which is characteristic of jet-style drift chambers in
  common-stop mode.}
\label{fig:church}
\end{figure}

Signals in the sense wires were processed with front-end electronics
housed in the recesses of the interconnecting sections before being
sent to TDC modules in the electronics hut. The signals were first
decoupled from the high-voltage on new, custom-designed, high-voltage
distribution boards. The signals next passed to Nanometrics
Systems\footnote{Nanometric Systems, Berwyn, IL, USA} N-277L
amplifier/discriminators. Then the signals were passed by Ethernet
cable to the electronics hut, to LeCroy\footnote{Teledyne Lecroy,
  Chestnut Ridge, NY, USA} 1877 Multihit TDC modules, operated in
common-stop mode, with the stop signal being provided by a delayed
trigger signal. The digitized signals were read out by the data
acquisition system. An example TDC spectrum for a single wire is shown
in Fig.~\ref{fig:church}.

\subsection{Time-of-Flight Detectors}
\label{sec:tof}
 
\begin{figure}[htbp] 
\centering 
\includegraphics[width=\columnwidth]{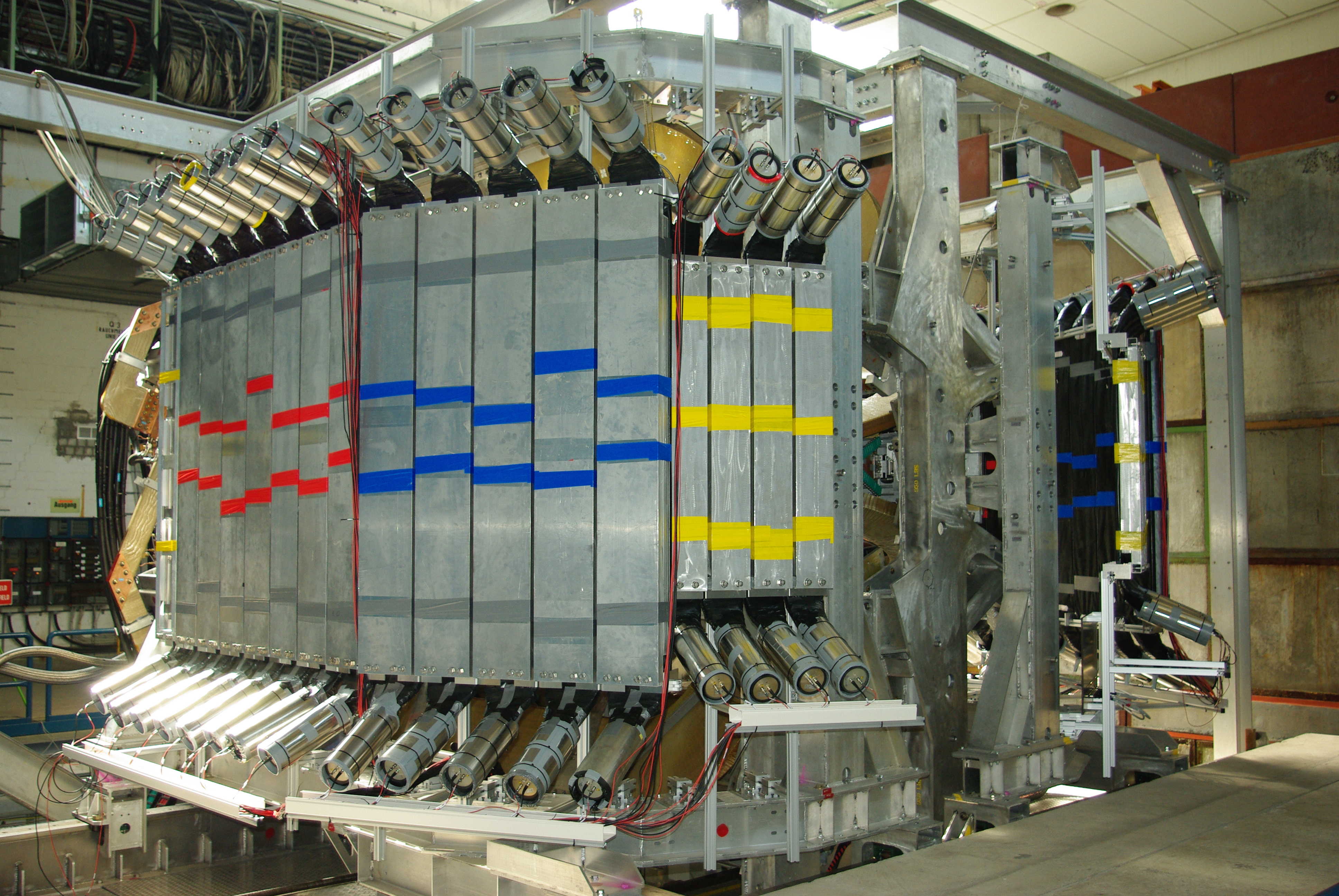}
\caption{Photograph of the mounted ToF detectors during assembly of
  the OLYMPUS detector.}
\label{fig:tof}
\end{figure}

The time-of-flight (ToF) detector was adapted from the system used for
the BLAST experiment~\citep{Hasell:2009zza}. Each sector consisted of
18~vertical scintillator bars read out with photo-multiplier tubes
(PMT) mounted at both ends, as shown in Fig.~\ref{fig:tof}.
 
The four most-forward bars on each side were 119.4~cm high, 17.8~cm
wide, and 2.54~cm thick. The remaining 14~bars on each side were
180.0~cm high, 26.2~cm wide, and 2.54~cm thick, so as to cover the
entire acceptance of the drift chambers.  The Glasgow University group
designed and constructed a new support structure which allowed a tight
arrangement and quick replacement of individual bars.  The bars were
arranged in three planar sections oriented with their normal
approximately pointing toward the target area. The rearmost two bars
in each sector were not present in BLAST and were added to expand the
acceptance of OLYMPUS at large~$\theta$.

The ToF detector provided the timing signals used to trigger the
readout and data acquisition system for the majority of detector
components. In particular, it provided the common-stop signal for the
drift chamber TDCs.  The main trigger logic of the experiment required
the presence of at least one top/bottom coincidence in both sectors
(see Sec.~\ref{sec:trigger}).  The ToF PMT signals were processed through
passive splitters and recorded by both TDCs and ADCs. The analog PMT
signals were discriminated with constant fraction discriminators for
the forward 16 bars on each side, and with leading-edge discriminators
for the rearmost two bars.  The logic signals were further processed
for the trigger which in turn provided the common-start signal for the
ToF TDCs and the common-stop signal for the drift chamber TDCs.  The
differential splitter outputs were connected to integrating ADCs.  The
integrated signal from a given bar provided an estimate of the energy
deposited in the bar, while the relative time difference between the
top and bottom tube signals from a bar provided a rough measurement of
the hit position. The mean signal times of the top and bottom signals
were approximately independent of the hit position. The difference in
mean times between pairs of ToF bars in opposite sectors measured the
difference in time-of-flight between scattered and recoiling particles
for interactions originating in the target or measured the
time-of-flight of cosmic ray particles traversing the detector.

The active volume of the ToF bars consisted of Bicron\footnote{Bicron,
  Solon, OH, USA} BC-408 plastic scintillator, chosen for its fast
response time (0.9~ns rise time) and long attenuation length
(210~cm). At the ends of each bar, the sensitive volumes were
connected via Lucite light guides to 3-inch diameter Electron
Tubes\footnote{Electron Tubes Ltd, Ruislip, Middlesex, England} model
9822B02 photomultiplier tubes equipped with Electron Tubes EBA-01
bases.  The PMT signals exhibited a typical amplitude of about 0.8~V
with a rise time of a few nanoseconds.  The light guides were bent
away from the interaction region to orient the PMTs roughly
perpendicular to the toroidal magnetic field.  Additionally, each PMT
was encased with mu-metal shielding. Due to these measures, the
toroidal magnetic field had no discernible effect on the ToF gains.
Each PMT base utilized actively-stabilized voltage dividers to avoid
variation of signal timing with gain.

Due to aging and radiation damage, some of the scintillator bars were
found to have short attenuation lengths.  This was determined by
examining the TDC and ADC signals for each bar.  Problematic bars were
replaced before data taking.

\begin{figure}[htbp] 
\centering 
\includegraphics[width=\columnwidth]{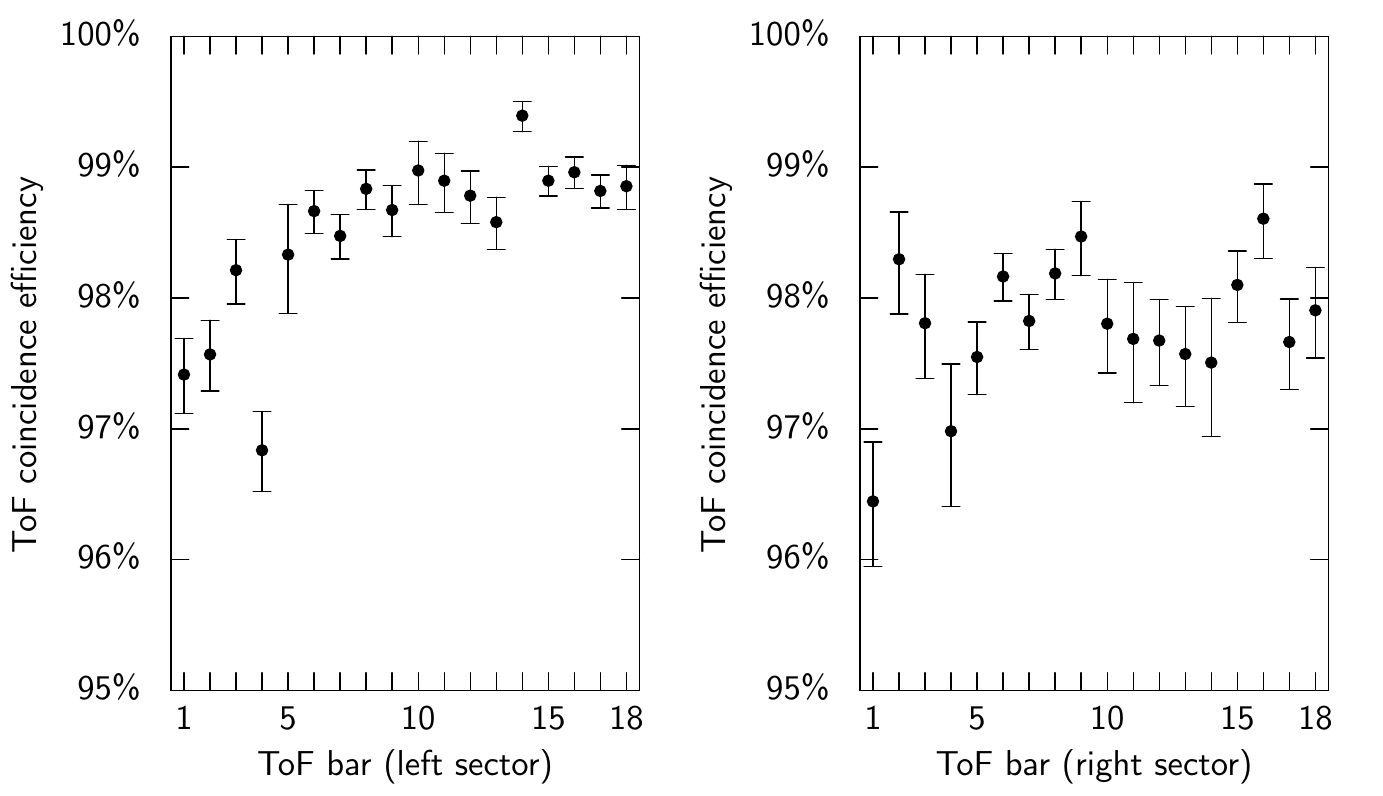}
\caption{Efficiencies for each TOF scintillator bar determined during
  the cosmic running period.}
\label{fig:tof_eff}
\end{figure}

After the experiment, during the cosmic ray runs, the efficiencies for
top/bottom coincidences were measured by sandwiching the center region
of each bar with a pair of small test scintillators.  These tests
found efficiencies to be around 96-99\% for signals originating near
the center of each bar as shown in Fig.~\ref{fig:tof_eff}.

\section{Luminosity Monitors}
\label{sec:lumi}

The physics goals of OLYMPUS required the very precise and accurate
measurement of the ratio of the integrated luminosities with positron
and electron beams delivered to the experiment. To achieve this,
OLYMPUS included three systems to measure the luminosity redundantly:
\begin{enumerate}
\item[-] The slow control system (Sec.~\ref{sec:slow}) monitored the beam
  current and gas flow to the target.  The system additionally used
  measurements of the target cell temperature, in conjunction with the
  known cell geometry, to compute the target density and thickness
  during running.  The product of the target thickness and beam
  current was corrected for the deadtime of the data acquisition
  system to produce a first estimate of the instantaneous luminosity.
\item[-] The $12\degree$ luminosity monitors (Sec.~\ref{sec:12deg})
  measured elastically scattered leptons in a small angular range in
  coincidence with the recoil proton detected in the opposite sector
  drift chamber.  Each monitor consisted of a telescope of three gas
  electron multiplier (GEM) detectors (Sec.~\ref{sec:gem}) interleaved
  with three multi-wire proportional chambers (MWPCs)
  (Sec.~\ref{sec:mwpc}). At $\theta = 12\degree$ the two-photon
  contribution to elastic scattering is expected to be negligible, the
  known $ep$ elastic cross section can be used to provide a luminosity
  measurement.  The system was designed to provide a luminosity
  measurement with a statistical precision better than 1\% each hour.
\item[-] A high precision measurement using symmetric M{\o}ller and
  Bhabha scattering was implemented using PbF$_2$ calorimeters placed
  symmetrically at $\theta =1.29\degree$ in the left and right sectors
  (Sec.~\ref{sec:moller}).  Comparing the observed $e^-e^-$ and $e^+e^-$
  elastic scattering rates with the known M{\o}ller and Bhabha cross
  sections provided a means of measuring of the luminosity for each
  beam species with very high statistical precision in very short time
  frames.
\end{enumerate}

\begin{figure}[htbp]
\centering
\includegraphics[width=\columnwidth]{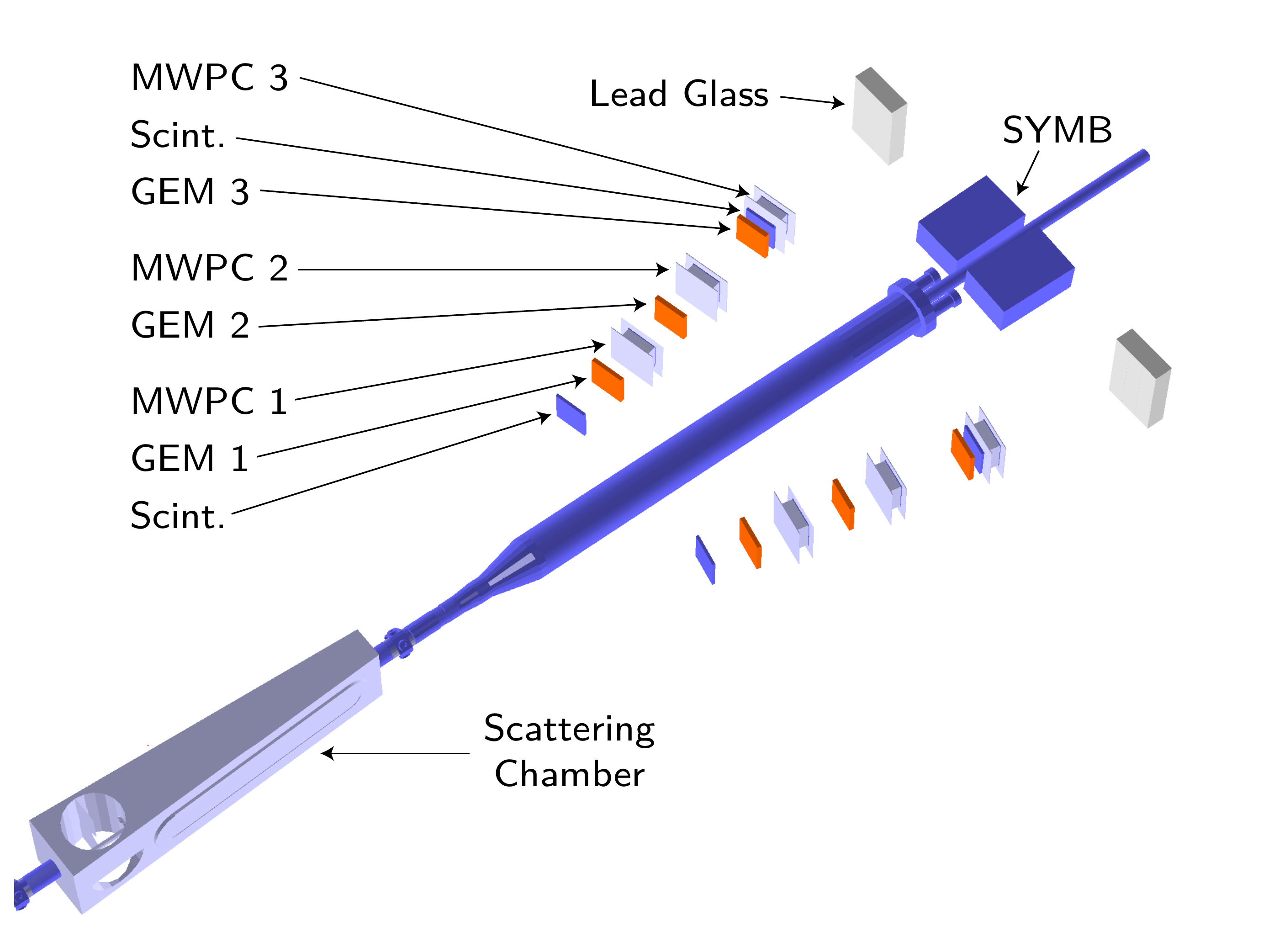}
\caption{Layout of the $\theta=12\degree$ luminosity monitors and the
  symmetric M{\o}ller/Bhabha calorimeters}
\label{fig:12deglumi}
\end{figure}
Fig.~\ref{fig:12deglumi} provides a schematic overview of the $12\degree$
and symmetric M{\o}ller/Bhabha luminosity monitoring systems.

\subsection{The $12\degree$ Luminosity Monitoring System}
\label{sec:12deg}

\begin{figure}[htbp]
\centering
\includegraphics[width=\columnwidth]{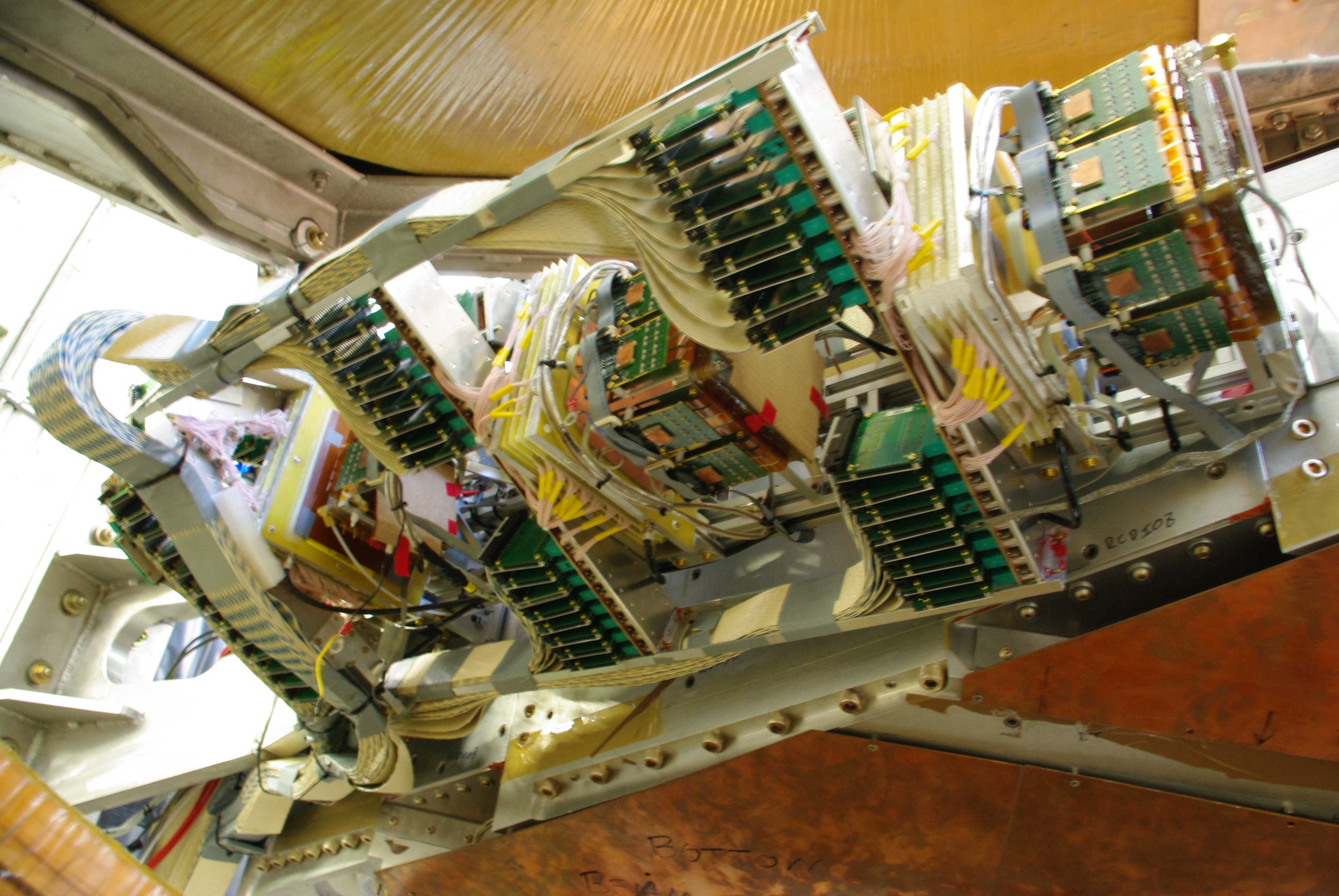}
\caption{Photograph of one of the $12\degree$ GEM/MWPC telescopes.}
\label{fig:12degphoto}
\end{figure}

The $12\degree$ luminosity monitoring system consisted of two
telescopes, each composed of three GEM and three MWPC detectors.  A
pair of thin scintillators with silicon photomultiplier (SiPM) readout
contributed to the trigger. The telescopes tracked leptons scattering
through small angles, a region where the asymmetry between electron
and positron scattering was expected to be small. The telescopes were
mounted to rails on the forward faces of the drift chambers to fit in
the space between the toroid coils on each side of the beamline. In
this position, the telescopes had a clear view of most of the target
cell. The two types of detectors provided redundancy for a high
efficiency measurement as well as a cross check against systematic
effects. A photograph of one of the $12\degree$ telescopes is shown in
Fig.~\ref{fig:12degphoto}.

\subsubsection{$12\degree$ GEM Detectors}
\label{sec:gem}

\begin{figure}[htbp]
\centering
\includegraphics[width=\columnwidth]{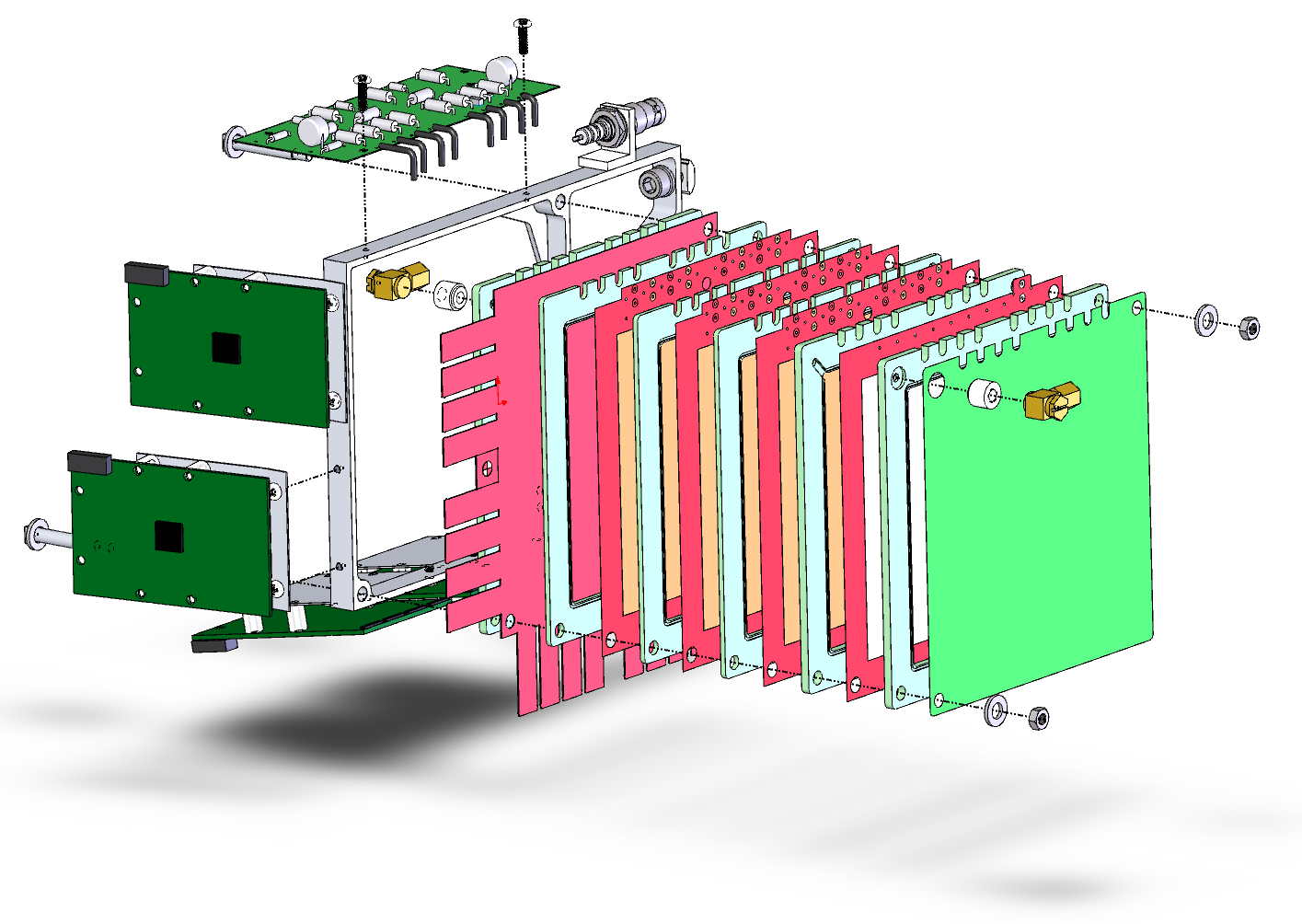}
\caption{An exploded view of a single triple-GEM detector.}
\label{fig:gem_exp}
\end{figure}

The triple-GEM detectors with 2D strip readout were designed at the
MIT-Bates Linear Accelerator Center and were constructed at Hampton
University.  INFN Rome provided the front-end and readout electronics
for the GEMs, which were designed in collaboration with INFN Genoa.
Each individual GEM chamber was constructed as a stack of frames and
foils glued together (see Fig.~\ref{fig:gem_exp}).
Each stack included a readout board with three GEM foils and a cathode
foil.  Two pressure volume foils enclosed the gas volume to avoid
deforming the readout or cathode foils had these been used to enclose
the gas volume.  There was a $2$~mm space between each GEM foil and
between the last GEM foil and the readout board.  The pressure volume
foils and the cathode foils were spaced $3$~mm from the adjacent
foils.  All of the components were tested individually before they
were assembled into a detector.  All of the electrical and gas
connections were accessible on the edges of the stack, or in special
cutouts in the case of the high voltage connections. A simple
resistive voltage divider card provided the high voltage to all
foils. A premixed, Ar:CO$_2$ 70:30 gas mixture was used.

The GEM, cathode, and readout foils were manufactured by
TechEtch\footnote{TechEtch Inc.  Plymouth, MA}.  Each GEM foil
consisted of 50 $\mu$m-thick Kapton clad on both sides with
5~$\mu$m-thick layers of copper. The GEM foils were chemically etched
to produce 70~$\mu$m holes in an equilateral triangular pattern with
140~$\mu$m pitch over the active area of the detector (approximately
10~cm~$\times$~10~cm).  The cathode foil consisted of 50 $\mu$m-thick
Kapton clad on only one side with a 5 $\mu$m-thick layer of copper and
no holes.  The cathode foil provided a uniform electric field
throughout the primary ionization area.  The pressure volume foils
consisted of 50 $\mu$m-thick aluminized Mylar, which additionally
served to electrically shield the detector.  The readout foil
consisted of a 50 $\mu$m-thick Kapton substrate foil.  On the charge
collection side there was precisely spaced pattern of lines and pads
of 0.5-1.0 oz.\@ (18-35 $\mu$m) gold-plated copper.  The lines aligned
vertically measured the horizontal coordinate of a hit.  Between each
pair of vertical lines there was a column of pads. Each pad was
connected with a via to the backside of the foil where they were
connected in horizontal rows to measure the vertical coordinate of a
hit.  The lines were 124~$\mu$m wide, at a 400~$\mu$m horizontal
pitch.  The pads were 124~$\mu$m~$\times$~323~$\mu$m, at a 400~$\mu$m
horizontal and vertical pitch. This geometry was chosen such that the
charge collected would be approximately equally shared between the
horizontal and vertical readout channels.

The signals from the lines and pads were routed to two edges of the
readout foil where they terminated on sixteen arrays of pads designed
to fit a flexible circuit connector, which was mounted on the
front-end electronics card.  Each card had four connectors (two cards
per coordinate) corresponding to a total of four cards per GEM
detector.  Each GEM detector had 500 channels (250 per coordinate),
with a total of 3000 readout channels for the GEMs in both telescopes.
The front-end readout card used one APV25-S1 analog pipeline chip per
card~\citep{French:2001xb}.  Each chip had 128 channels, each of which
had a 192-cell analog pipeline which sampled the input channels at
40~MHz.  Data were read out of the pipeline after a trigger event.
All 128 channels were multiplexed onto a single data line read out by
the DAQ system.  The communication between the APV card and the DAQ
system was maintained by the Multi-Purpose Digitizer
(MPD)~\citep{Musico:2011lia}.  The MPD consisted of a VME-based module
that hosted digital bus drivers, fast ADCs, and a field-programmable
gate array (FPGA).  The FPGA was responsible for the configuration,
synchronization, triggering, and digitization of the APV cards and the
data transfer along the VME bus.

The GEM detectors were fixed to an aluminum mounting bracket attached
to the rails that also held the MWPCs.  The mounting bracket had
flexible supports for the high voltage and front-end electronics
cards.  These allowed the positions of the cards to be adjusted during
installation to avoid interference between components.  Both the
mounting bracket and rails were adjustable.  Survey targets located on
the GEM chambers allowed the detector positions to be measured.

A charged particle traversing the GEM elements produced a charge
cluster which was registered by several lines and pads in both the
vertical and horizontal directions.  The reconstructed centroid of the
clusters in $x$ and $y$ gave the spatial location of the particle as
it passed through the detector. Digitization of the signal amplitudes
of all channels allowed the detector to achieve high spatial
resolution (70~$\mu$m). The efficiency of each GEM detector was
measured with candidate tracks based on the other five telescope
elements and was found to be around 95\% for all GEM elements.

\subsubsection{$12\degree$ Multi-Wire Proportional Chambers}
\label{sec:mwpc}

\begin{figure}
\centering
\includegraphics[width=\columnwidth]{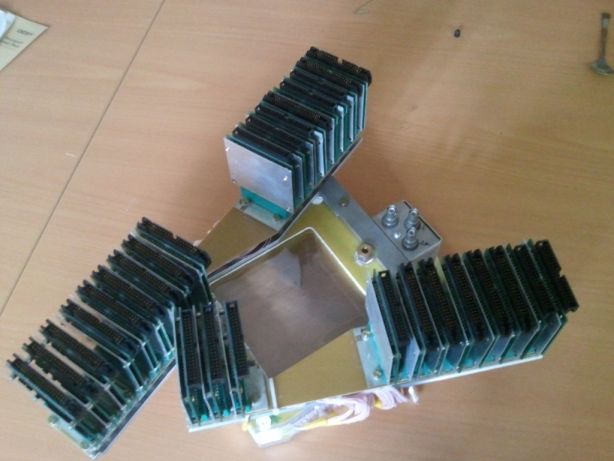}
\caption{Photograph of one MWPC with CROS3 readout electronics.}
\label{fig:mwpc2}
\end{figure}

Six identical MWPC modules, along with their CROS3 readout
electronics~\citep{Uvarov:cros3}, were fabricated at PNPI for the
$12\degree$ luminosity telescopes.  Each MWPC module had the external
dimensions 180~mm~$\times$~180~mm~$\times$~50~mm and an active area of
112~mm~$\times$~112~mm. The readout cards for each module were
arranged in two stacks around the active area, as shown in
Fig.~\ref{fig:mwpc2}, to fit in the narrow space between the toroid coils.
 
Each MWPC module consisted of three planes of anode sense wires
interleaved with cathode wire planes. The gap between the anode and
cathode planes was 2.5~mm. The anode sense wires were angled with
respect to each other to allow hit reconstruction in two dimensions (X
vertical, $0\degree$, U $+30\degree$, and V $-30\degree$). The sense
wires were 25~$\mu$m-diameter, gold-plated tungsten separated by
1~mm. The cathode wires were 90~$\mu$m-diameter beryllium bronze
separated by 0.5~mm. Each plane of wires had its own fiberglass
frame. The module was assembled by sandwiching the planes together in
a 10~mm aluminum outer frame. Each MWPC detector had material
thickness of 0.25\%~X$_0$ in the active area.

\begin{figure}[htbp]
\centering
\subfloat[X plane]{
  \includegraphics[width=0.33\columnwidth]{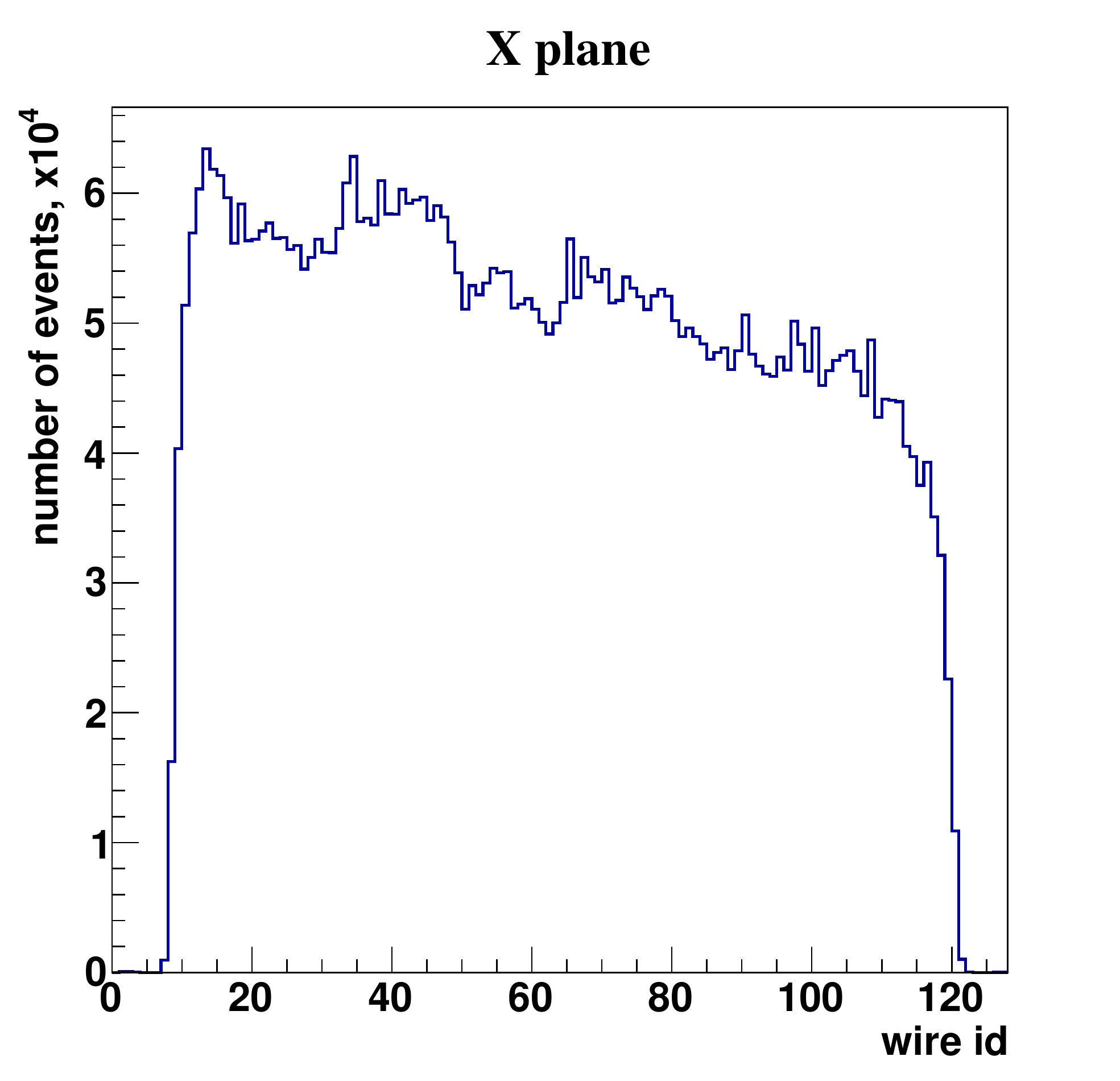}}
\subfloat[U plane]{
  \includegraphics[width=0.33\columnwidth]{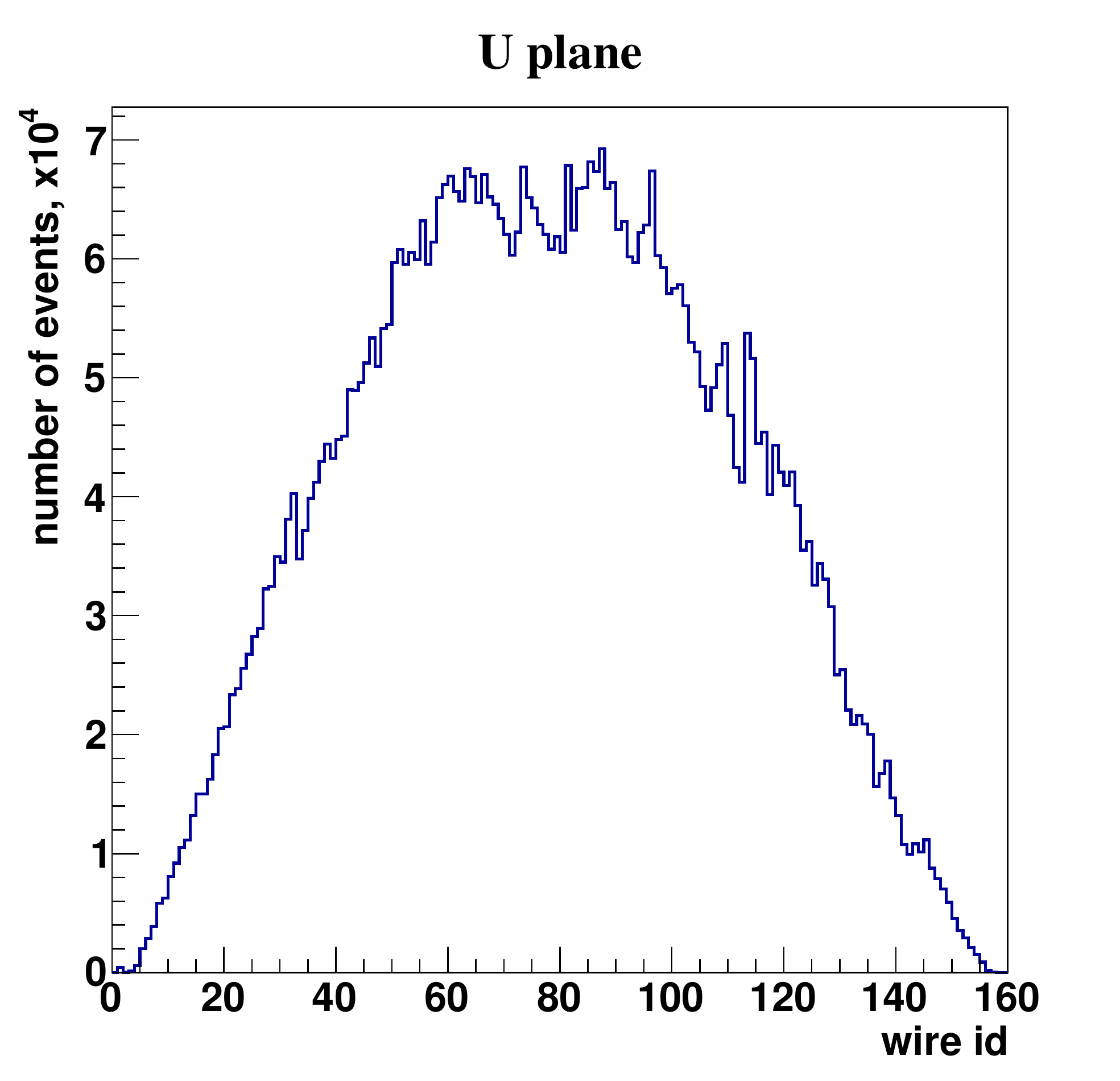}} 
\subfloat[V plane]{
  \includegraphics[width=0.33\columnwidth]{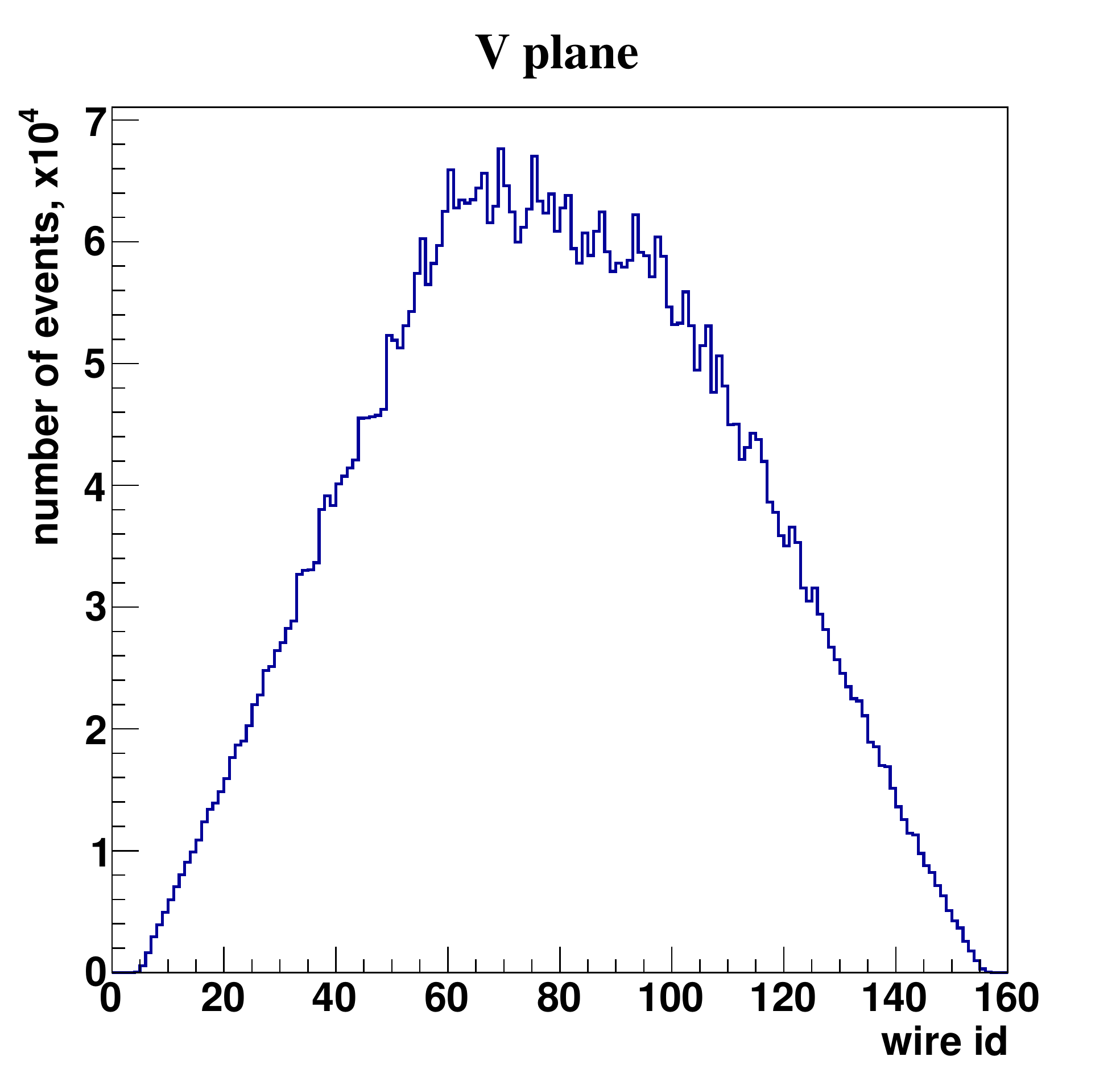}}
\caption{Hit distributions for the X, U, and V planes of a single,
  representative MWPC detector. The distribution of events is
  determined by the detector acceptance and sense wire angle for each
  plane.}
\label{fig:mwpc4}
\end{figure}

A gas mixture of Ar:CO$_2$:CF$_4$ (65:30:5) was chosen for the MWPCs
based on the experience gained from the proportional chambers produced
at PNPI for the HERMES experiment~\citep{Andreev:2001kr}. GARFIELD
\citep{Veenhof:1998tt} calculations predicted a gas gain of
$7\times10^4$ in the MWPCs at 3150~V. During operation, 3200~V was
used after testing the MWPCs with a $^{55}$Fe radioactive source.
This operating voltage was validated during running, where an
efficiency of 98--99\% was typically seen for all MWPC modules.  Hit
distributions for each plane in a single MWPC detector are presented
in Fig.~\ref{fig:mwpc4}.

\subsubsection{$12\degree$ Trigger}
\label{sec:SiPM}

Each $12\degree$ telescope included two
120~mm~$\times$~120~mm~$\times$~4~mm scintillator tiles
(Eljen\footnote{Eljen Technology, Sweetwater, TX, USA} EJ-204) to
provide a trigger signal for the GEMs and MWPCs.  Each scintillator
tile was wrapped in diffuse reflectors (Millipore\footnote{EMD
  Millipore, Billerica, MA, USA} Immobilon-P) and read out using two
Hamamatsu\footnote{Hamamatsu Photonics K.K. Hamamatsu, Japan} SiPM
multi-pixel photon counters (MPPCs) mounted on two opposing
corners. This ensured a very high homogeneity of the light yield from
the entire area of the tiles.  The analog signals from each MPPC were
summed and constant fraction discriminators provided the output signal
from each tile.  The trigger for reading out the $12\degree$ telescope
on a given side consisted of the triple coincidence of the two tiles
on that side in conjunction with a trigger from a ToF bar in the rear
region of the opposite side of the detector.

Additionally, lead glass calorimeters mounted behind the $12\degree$
telescopes in each section provided an independent means of triggering
the detectors.  Each calorimeter consisted of three lead glass bars
attached to a PMT for readout.  The additional trigger contributed the
ability to measure the efficiency of the tile trigger continuously
throughout data taking.  The two scintillator tiles in each telescope
exhibited combined efficiencies in excess of 99\% throughout the
experimental run.

\subsection{Symmetric M{\o}ller/Bhabha Luminosity Monitor}
\label{sec:moller}

\begin{figure}[htbp]
\centering
\includegraphics[width=\columnwidth]{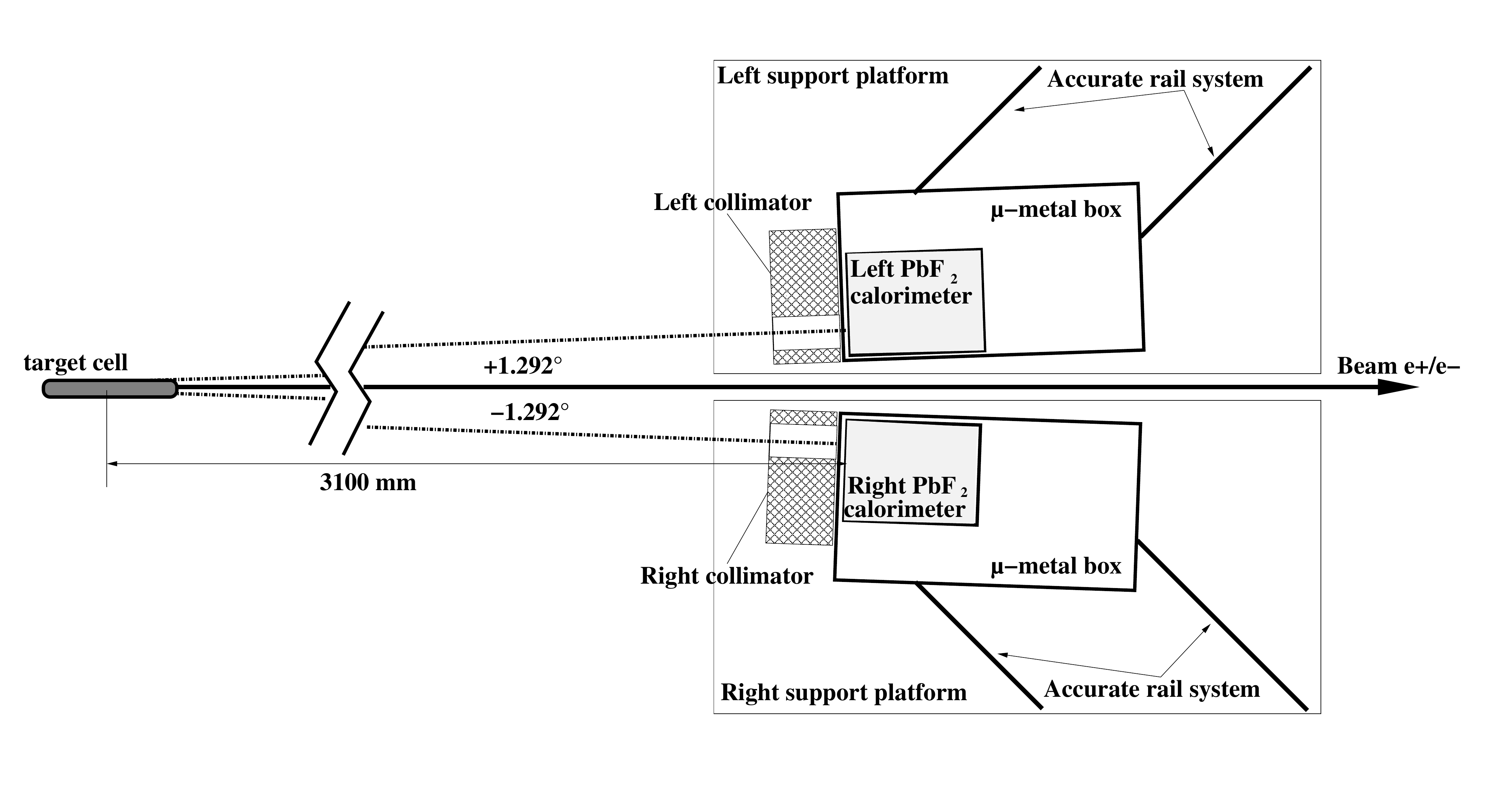}
\caption{A schematic of the Symmetric M{\o}ller/Bhabha luminosity
  detector (SYMB) showing the symmetric design about the beamline.}
\label{fig:draw_schema}
\end{figure}

\begin{figure}[htbp]
\centering
\includegraphics[width=\columnwidth]{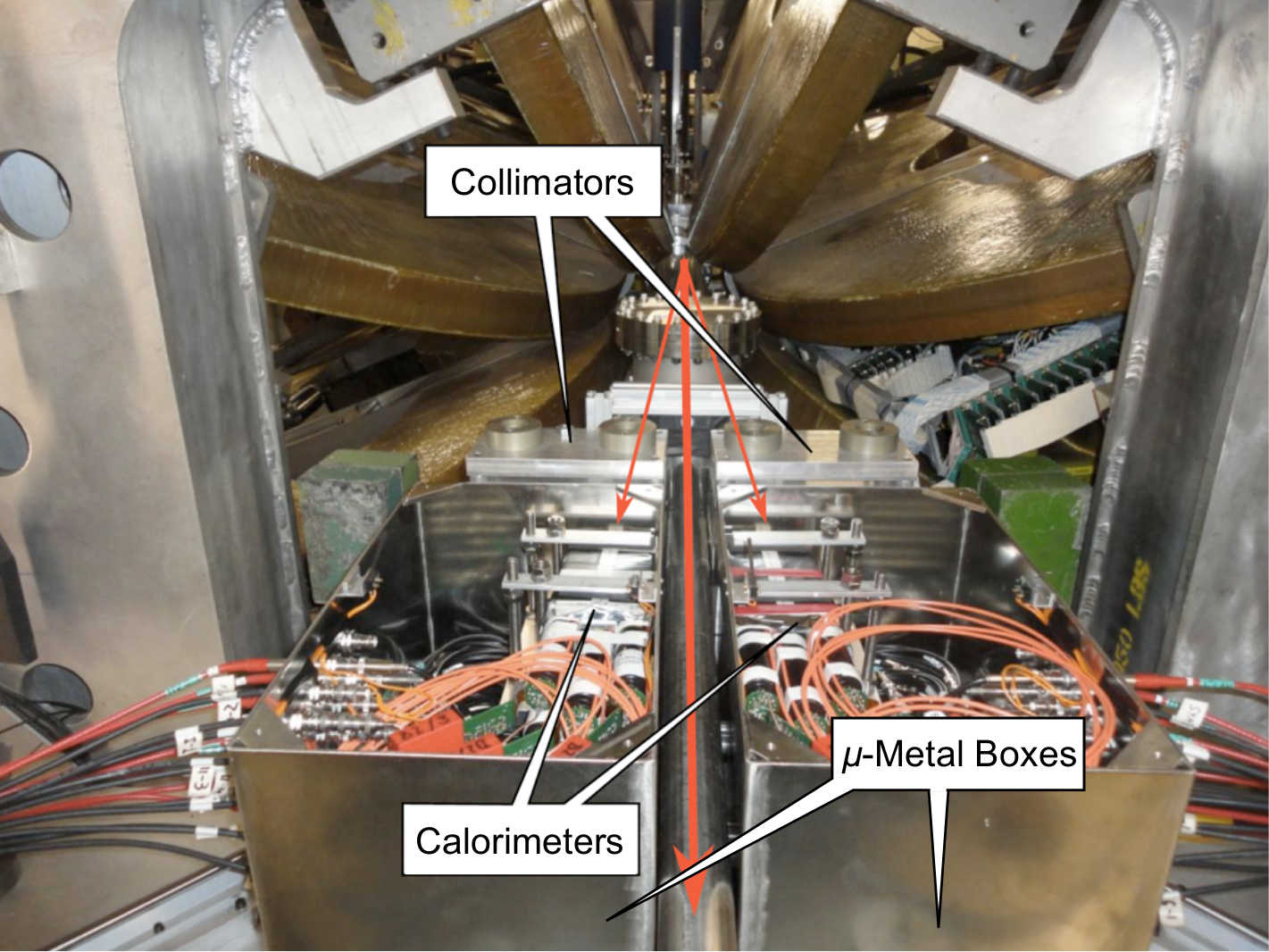}
\caption{A photograph showing the main components of the SYMB
  detector.  The thick red line indicates the direction of the beam
  while the thinner red lines indicate the general path of scattered
  electrons, positrons, or photons entering the SYMB.}
\label{fig:photo}
\end{figure}

The symmetric M{\o}ller/Bhabha (SYMB) calorimeter measured the
coincidence rate of lepton-lepton scattering events at symmetric
angles. The cross sections for these processes are precisely
calculable from quantum electrodynamics, and the rates in the SYMB
were high enough to yield an luminosity measurement on the timescale
of minutes. During electron beam running, the detector recorded
M{\o}ller scattering events ($e^{-}e^{-}\rightarrow e^{-}e^{-}$),
while during positron running it was sensitive to both Bhabha
scattering ($e^{+}e^{-}\rightarrow e^{+}e^{-}$) and annihilation
($e^{+}e^{-}\rightarrow \gamma \gamma$) events. At the OLYMPUS beam
energy of 2.01~GeV, symmetric scattering occurred at a polar angle of
$1.292\degree$ with respect to the beam direction
(see~Figs.~\ref{fig:draw_schema} and~\ref{fig:photo}).

\begin{figure}[htbp]
\centering
\includegraphics[width=0.49\columnwidth]{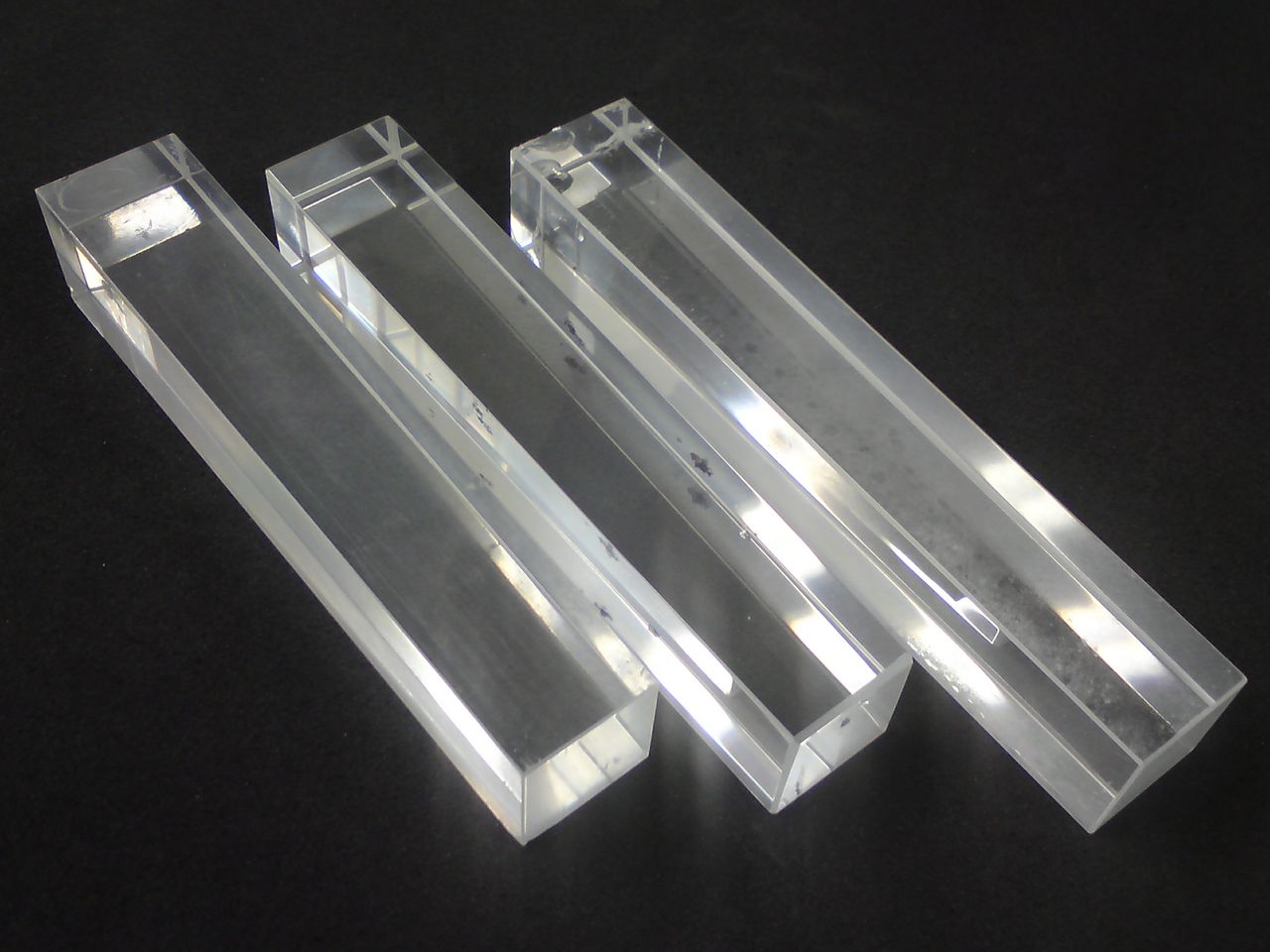}
\includegraphics[width=0.49\columnwidth]{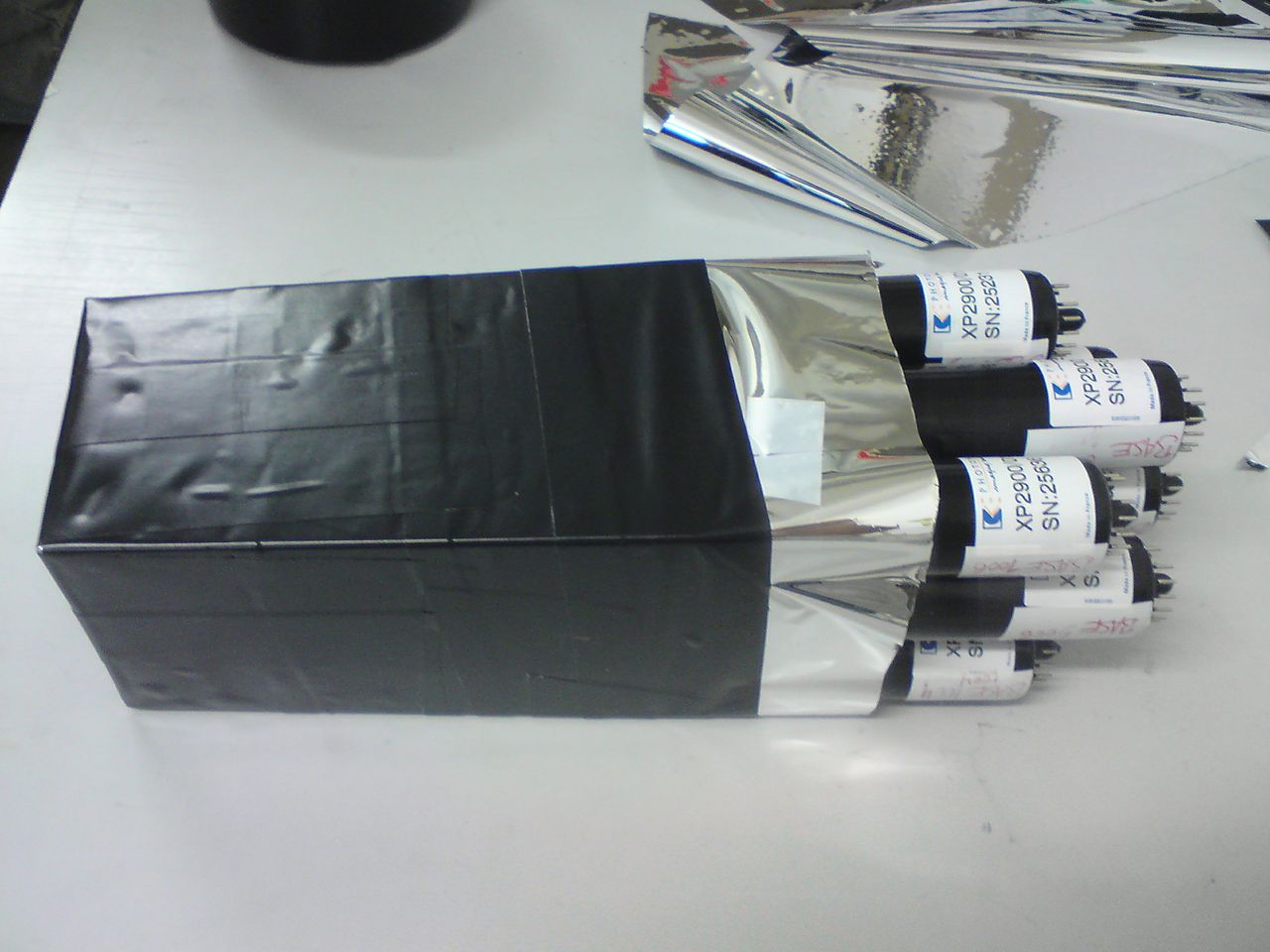}
\caption{Several of the PbF$_2$ crystals used in symmetric
  M{\o}ller/Bhabha luminosity monitor before (left) and after (right)
  assembly with the PMT readout system.}
\label{fig:SYMB}
\end{figure}

The SYMB, constructed at Johannes Gutenberg-Universit\"at in Mainz,
Germany, consisted of two $3\times3$ arrays of lead fluoride
(PbF$_{2}$) crystals, as shown in~Fig.~\ref{fig:SYMB}.
A Philips\footnote{Koninklijke Philips N.V., Amsterdam, the
  Netherlands} XP 29000/01 PMT was connected to the end of each
crystal to provide readout.  The SYMB was able to operate at high
rates because of the fast response of the PMTs (20~ns), and because
showers in PbF$_2$ produce only \v{C}erenkov radiation, eliminating
the delay associated with a scintillation signal. Each crystal was
approximately $26$~mm~$\times$~$26$~mm~$\times$~$160$~mm, with a
slightly tapered shape.  An array of crystals was more than 15
radiation lengths long and extended approximately 2~Moli\`ere radii
from the center to the nearest edge~\citep{Baunack:2011pb}.  Millipore
paper wrapping around each crystal increased the surface reflectivity
to reduce light loss.  Each array of crystals and PMTs resided inside
a mu-metal box to shield them from the magnetic fields of the OLYMPUS
toroid and the DORIS beamline quadrupoles.

Lead collimators, located between each detector array and the target,
shielded the crystals from beam bremsstrahlung, non-symmetric
M{\o}ller/Bhabha events, and other backgrounds.  Each collimator
consisted of a 100~mm thick lead block with a precision-machined
circular hole with a diameter of 20.5~mm.  Since these apertures
determined the solid angle acceptance of each detector, the location
and orientation of the collimator holes was carefully surveyed before
and after each running period.

The SYMB readout electronics were based on a design used for the A4
experiment at MAMI in Mainz~\citep{Kothe:2008zz}.  The system provided
the ability to conduct fast analog summation of the nine PMT signals
from each crystal array and to quickly digitize and histogram the
summed signal.  The detector operated up to a rate of 50~MHz (limited
by the 20~ns signal time of the PMTs).  Since the minimum bunch
spacing during OLYMPUS operation was 96~ns there was no deadtime
associated with the SYMB readout.  Typical single event rates were
15~kHz, well within the operational capabilities of the readout
electronics. The SYMB electronics were suspended while the OLYMPUS
readout system (Sec.~\ref{sec:readout}) processed an event to ensure the
SYMB counted events only while the trigger was open.

A crystal array generated a trigger signal if two conditions were
met. The first condition required that the sum of analog signals in
all nine crystals exceeded the threshold of a constant fraction
discriminator. The second condition required that the central crystal
have the largest signal, in order to reject noise events. Upon
receiving a trigger signal, the detector electronics would histogram
the event. One histogram was for events when both arrays produced a
trigger. Two additional histograms were filled when the left or right
arrays respectively produced a trigger. Due to the high event rate,
single events were not read out. Rather, the histograms were
periodically sent to the data acquisition system.

\begin{figure}[htbp]
\centering
\includegraphics[width=\columnwidth]{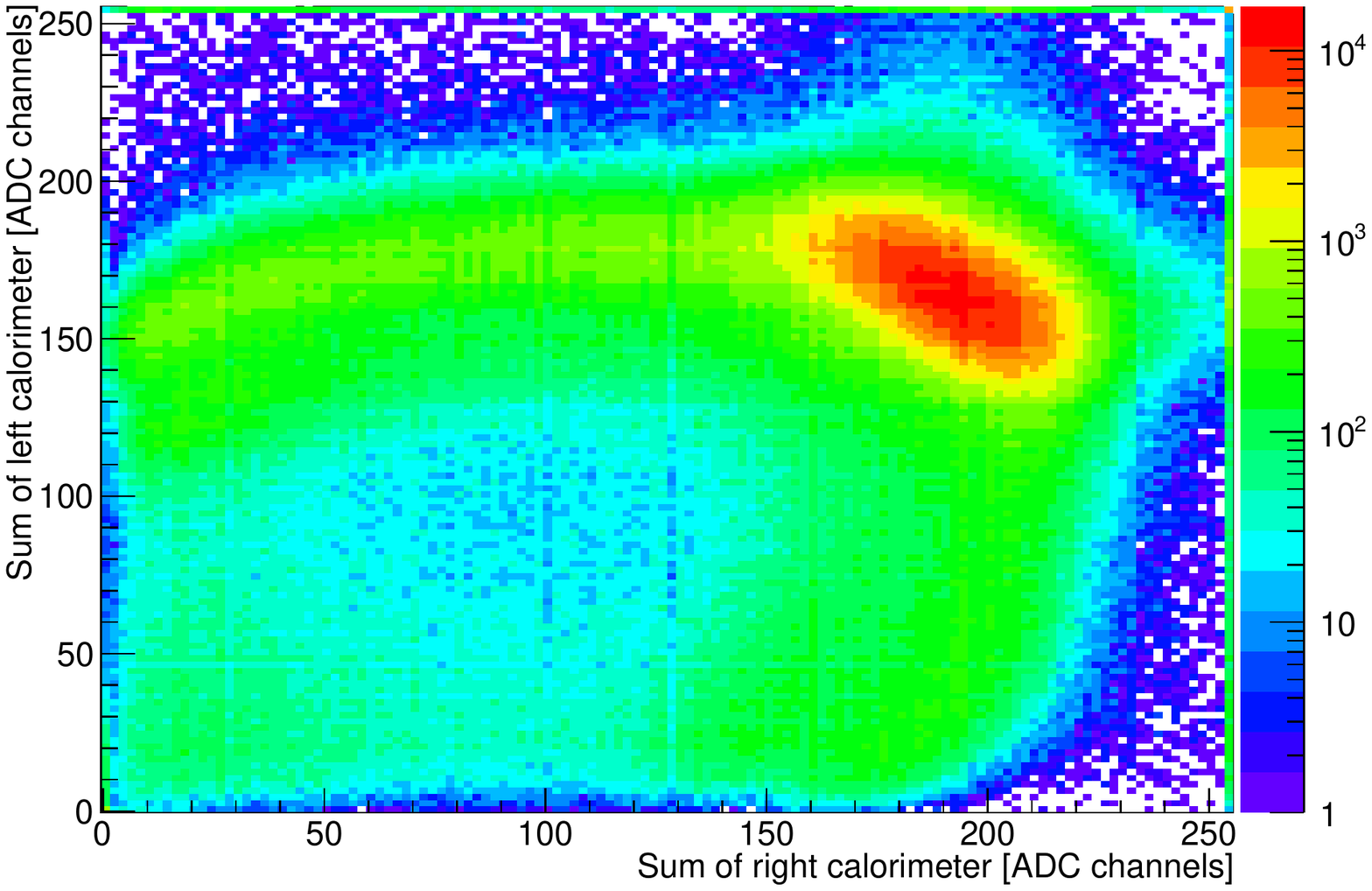}
\caption{A 2D histogram of the sum of the deposited energy in the left
  and right SYMB calorimeters in coincidence mode. }
\label{coincidence}
\end{figure}

Fig.~\ref{coincidence}
shows an example of the coincidence-event histogramming.  Symmetric
M{\o}ller, Bhabha, and annihilation events deposit approximately the
same energy in both calorimeters, while many background processes
deposited energy asymmetrically.

\section{Data Acquisition}
\label{sec:DAQ}

The data readout and trigger system for OLYMPUS was developed in
collaboration between the Bonn and MIT groups, based on the system
originally developed for the Crystal Barrel experiment
\citep{CrystalBarrel} at ELSA in Bonn, Germany. The system employed
VME CPUs, using standard 1~GBit Ethernet for data transport, and
dedicated hardware connections and modules for
synchronization. Trigger logic was implemented by a flexible FPGA
system.  The data acquisition system was controlled through a
graphical user interface.  The following two subsections describe
these systems in more detail.

\subsection{Trigger}
\label{sec:trigger}

The OLYMPUS trigger system incorporated information from the
time-of-flight detector, the drift chambers, the luminosity detectors,
as well as information from the DORIS accelerator.  This was
implemented using a VME field-programmable gate array (FPGA), which
combined up to 16 input signals to produce 16 parallel trigger
conditions.  The individual conditions could be independently
prescaled.

The ToFs and the $12\degree$ scintillators provided the fast trigger
signals for the experiment. The primary trigger required a coincidence
between a left ToF bar and a right ToF bar that could be hit by a
kinematically valid elastic scattering event. Coincidence between the
top and bottom PMTs in each bar was required as well.  The main
$12\degree$ luminosity trigger consisted of a coincidence between the
two $12\degree$ scintillators in one sector in conjunction with a ToF
in the opposite sector. The DORIS bunch clock was used to provide the
reference time signal for the ToF and drift chamber TDCs.

In addition to the primary triggers, several signals corresponding to
less strict ToF coincidences and signals from the lead glass
calorimeters behind the $12\degree$ detectors were included at higher
prescale factors.  Events from these triggers provided means of
monitoring the efficiencies and calibration of various detector
components over the course of data taking.

The data from the February run contained an unsatisfactory fraction of
elastic $e^\pm p$ events. A second-level trigger that incorporated
information from the drift chambers was implemented for the fall run.
The trigger required a signal from at least one wire in each of the
middle and outer chambers on each sector and executed a fast clear of
the trigger when this condition was not satisfied. This scheme reduced
the false trigger rate by a factor of approximately 10.

\subsection{Readout}
\label{sec:readout}

The readout system was designed and implemented by the Bonn group,
based on VME CPU modules. The readout was designed in synchronous
fashion. An accepted trigger would cause all detectors to be read out,
while simultaneously inhibiting new triggers until the readout
procedure of all detectors was completed.  While a synchronous system
incurs a higher deadtime than an asynchronous system, the guaranteed
matching of data from different detectors for the same event and the
ease of identifying readout problems outweighed this disadvantage for
OLYMPUS.  The detector readouts were organized in a master-slave
architecture. Detectors were read out through a series of slave
modules with dedicated links to a master module, which sequenced the
readout. Upon receipt of a trigger, the master module would signal the
slave modules to begin readout and then wait until all slave modules
signaled that the procedure was completed.  These signals were
communicated over direct hardware lines while data transfer and
general control were facilitated by two dedicated 1~GBit Ethernet
networks.

\section{Slow Control}
\label{sec:slow}

The operation of the OLYMPUS experiment required several hundred
parameters to be monitored, controlled, and recorded.  These included
high voltage supplies, vacuum pumps and gauges, the hydrogen gas
supply system, the parameters of the DORIS beam, and other elements
with operational time scales on the order of seconds.  To satisfy
these requirements, a new dedicated slow control system was developed
for OLYMPUS.

The slow control system utilized the Experimental Physics and
Industrial Control System
(EPICS)\footnote{http://www.aps.anl.gov/epics/index.php} as its
backend solution.  The system ran on three Linux machines: two VME
computers with interface cards connecting to the control equipment and
one server which communicated data to a PostgreSQL database and
interfaced with the DORIS control system.  The database recorded the
status and history of all parameters associated with the slow control.
The slow control also passed this information to the DAQ for
integration with the detector data to produce the run data files.

The slow control system included a user-friendly, web-accessible
graphical user interface, implemented using
Flask\footnote{http:/flask.pocoo.org} as middleware.  While typical
slow control systems require the deployment of custom, operating
system dependent software on their control computers, the design of
the OLYMPUS system allowed both view-only and control access from any
computer with an Internet connection.  The user interface provided
simple on-screen controls for the various elements connected to the
system, displayed real-time plots and indicators of system statuses
and data, and produced visual and audible alarms when parameters
failed to satisfy proper run conditions.

\section{Operation}
\label{sec:operation}

\begin{figure}[htbp]
\centering 
\includegraphics[width=\columnwidth]{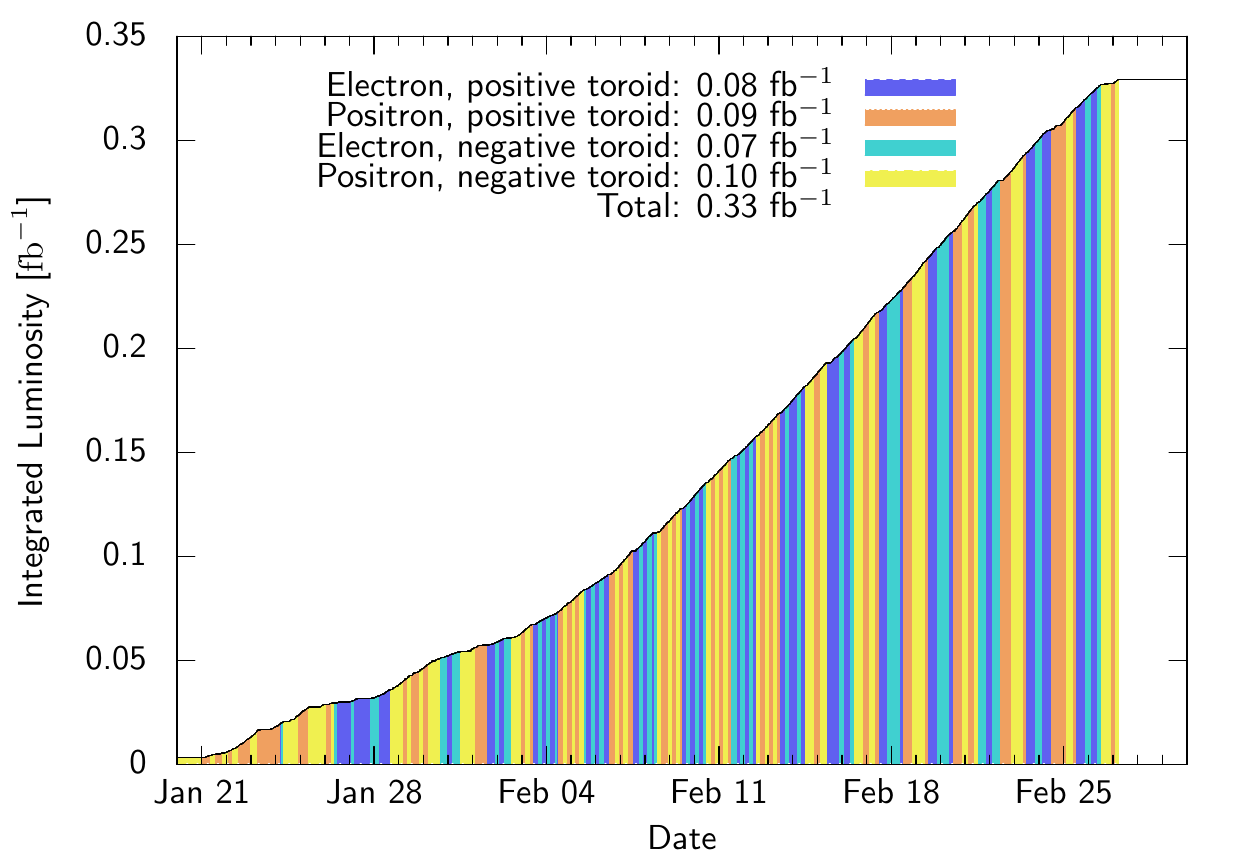}
\includegraphics[width=\columnwidth]{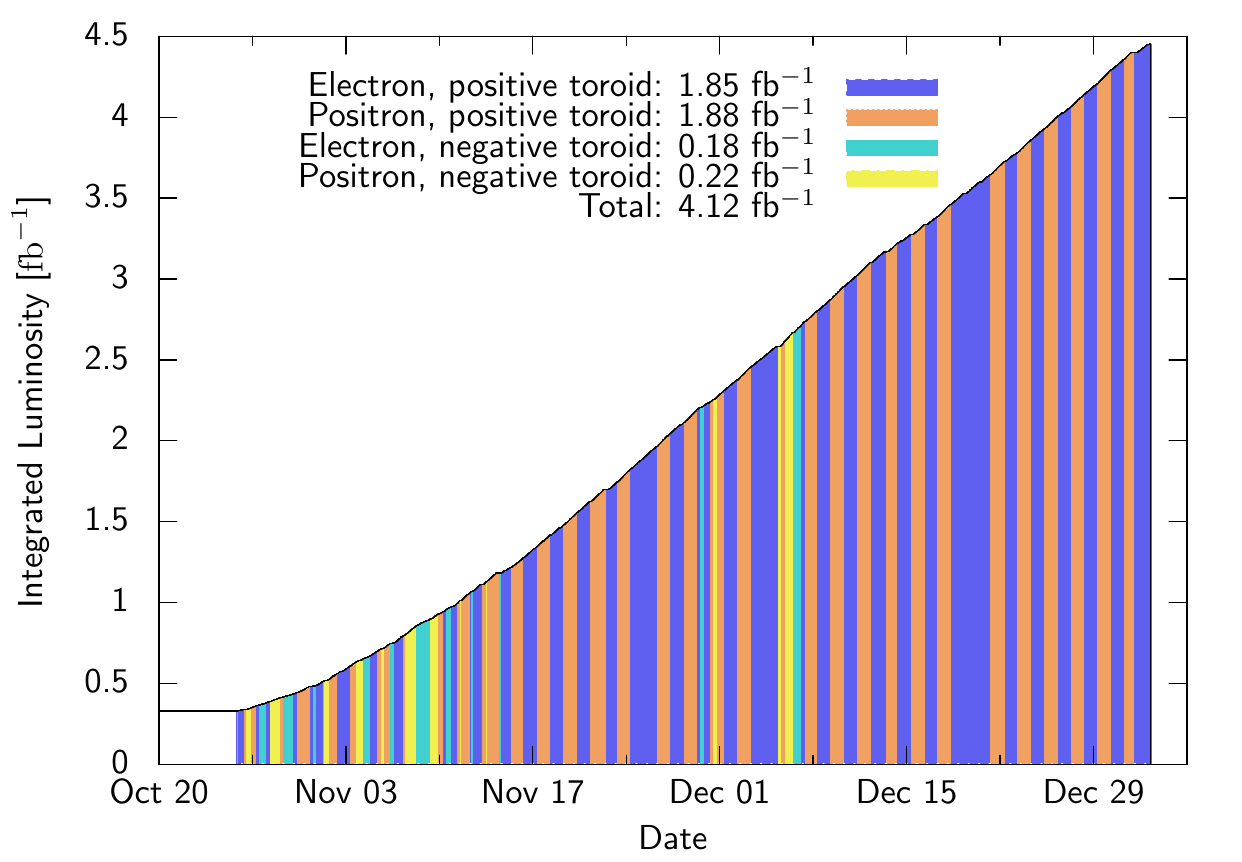}
\caption{The approximate integrated luminosity delivered to the
  OLYMPUS experiment during the February (left) and fall runs (right),
  as measured by the slow control (accurate to 10\%).}
\label{fig:intlumi}
\end{figure}

During normal data-taking runs, a two-person shift crew operated the
OLYMPUS detector and monitored the quality of the data using a number
of plots generated in near real-time.  Typically, production runs were
taken 24 hours a day during the February and fall runs, alternating
daily between positron and electrons beams.  The integrated luminosity
delivered to the experiment during the two runs is shown in
Fig.~\ref{fig:intlumi}.
In total, a data set of approximately 4.5~fb$^{-1}$ was collected over
the course of both runs. As discussed in Sec.~\ref{sec:intro}, the density
of gas in the target cell during the February run was significantly
lower than the design value because of a leak between the H$_2$ gas
feed system and the target cell.  Due to this, less than 10\% of the
ultimate data set was collected during the February run.  As is
described in the following section, it was possible to run at a higher
average beam current during the fall run, which allowed the experiment
to reach the design goal for the integrated luminosity.  At these
higher currents, however, it was difficult to operate the experiment
using a negative toroid polarity since low energy electrons were bent
into the detectors, resulting in a very high background level.  Thus,
negative polarity runs were only taken occasionally, with reduced beam
current and target flow. The uptime during the data-taking runs was
extremely high (approximately 95\%), with most of the downtime due to
the time required to switch the beam species (on the order of an
hour).

\subsection{Data Collection}

As noted in Sec.~\ref{sec:doris}, the experiment employed two modes of
operation, differentiated by the manner in which the DORIS ring was
operated.  During the February run, the experiment was operated in
``manual'' mode in which the beam was initially filled to 65~mA and
then data were taken until the beam decayed to 40~mA.  At this point,
the shift crew used the slow control interface (Sec.~\ref{sec:slow}) to
lower the high voltage of the various detectors to safe preset values.
Since beam refills during the earlier running period were not as clean
as during the fall 2012 run (more instability and losses), the
lowering of the voltages prevented high voltage trips and possible
damage to the detectors during the refill.  After lowering the
voltages, the OLYMPUS shift crew informed the DORIS accelerator crew
that the detector was ready for beam refill.  Once the beam was
restored to the normal starting current, the voltages were brought
back to operational values and data-taking was restarted.

Between the February and fall runs, significant improvements were made
to the DORIS beam injection process that allowed OLYMPUS to be run in
``top-up mode.''  In this mode, the beam was initially filled to 65~mA
as in the manual mode, but was only allowed to decay to 58~mA before
triggering an automatic refill.  Due to the improved injection, it was
not necessary to lower the high voltage of the OLYMPUS detectors
during these injections. The DAQ was configured to briefly inhibit
data-taking during injection pulses (see Sec.~\ref{sec:doris}).  This mode
of running significantly increased the average instantaneous
luminosity delivered to the experiment and freed the OLYMPUS shift
crew to more carefully monitor the quality of the incoming data.

The switch between beam species took place each morning, with
occasional exceptions for maintenance and balancing the amount of data
collected with each species. This ensured that there were no
systematic differences between $e^+$ and $e^-$ runs introduced by
environmental factors such as day/night cycles, reduced activity on
the DESY campus on weekends, etc.  During the February run, when both
toroid polarities were used, data-taking was segmented into four
six-hour blocks each day.  The pattern of toroid polarities in the
four blocks each days was selected by coin toss to ensure equal
running time for each polarity while avoiding systematic effects due
to the time of day and week.

In addition to production runs, empty target runs (with the H$_2$ gas
flow shut-off and the target chamber pumped down to ring vacuum
levels), zero magnetic field runs, and other test runs were taken on
an approximately daily basis for the purposes of monitoring
backgrounds, providing data for detector calibrations, and testing
proposed changes to operations.  When the DORIS beam was unavailable
due to problems or maintenance, the detector was left active to
collect cosmic ray data.  Also, cosmic ray data were collected for
approximately one month following the end of OLYMPUS production runs
in January 2013.  This large cosmic data set is being used for various
studies of detector efficiencies and for calibration.

\subsection{Data Quality Monitoring}

During data taking, the quality of the incoming data was monitored in
several stages.  Real-time, online monitoring of essential parameters
was implemented using the ExPlORA framework originally developed by
the Crystal Barrel collaboration~\citep{Piontek:2006ex}.  The ExPlORA
program processed the raw data files during data collection to produce
a variety of histograms and plots of quantities versus time, such as
the number of drift chamber wires hit per event, ADC and TDC
distributions, DAQ deadtime, and various detector rates.  The OLYMPUS
shift crew had access to reference plots corresponding to those shown
in ExPlORA that showed data of known good quality and data
representing known possible issues.  This provided the shift crew with
the ability to quickly identify problems with detectors as well as
problems caused by poor beam quality and take action to resolve them.

For the fall run, a second level of data quality monitoring by the
shift crew was implemented that allowed inspection of the data in a
more processed format approximately 30 minutes after the data was
taken.  This program automatically ran basic analysis programs on
complete datasets as they became available and presented the data to
the shift crew.  In a similar fashion as the real-time monitoring,
this program presented histograms and plots of the recent data to be
compared with data of known quality, but included higher-level
information such as the properties of events with good particle track
candidates and basic measures of detector efficiencies.

Additionally, the long-term performance of the detector was monitored
using the slow control database discussed in Sec.~\ref{sec:slow}.  This
provided the ability to monitor the behavior of many detector
parameters over the course of the entire data-taking period to
identify slow drifts and sudden changes that could affect the
analysis.

\section{Summary}
\label{sec:summary}

In 2012, the OLYMPUS experiment successfully collected approximately
4.5~fb$^{-1}$ of data for electron and positron elastic scattering
from hydrogen at the DORIS storage ring at DESY.  The experiment used
a large acceptance, left/right symmetric detector system consisting of
a toroidal magnetic spectrometer with drift chambers for tracking,
time-of-flight scintillators for triggering and relative timing, and a
redundant set of luminosity monitors.  A flexible trigger and data
acquisition system was used to collect the data. The left/right
symmetric design of the detector and the daily change of beam species
further reduced the systematic uncertainties of the measurement. The
initial plan to change the toroidal magnet polarity daily was not
possible due to high background rates in the negative polarity
configuration.  Consequently the majority (87\%) of the data were
collected with positive magnet polarity.

This paper has provided a technical description of the accelerator,
internal target, detectors, data acquisition, and operation of the
OLYMPUS experiment. Additional papers will detail the detector
performance, analysis, and physics results.

\section{Acknowledgments}
\label{sec:acknow}

The successful design, construction, and operation of the OLYMPUS
experiment would not have been possible without the research and
technical support staffs of all of the institutions involved.  In
particular, we would like to acknowledge the DORIS accelerator group
for providing the high quality electron and positron beams delivered
to the experiment.  We also gratefully acknowledge the DESY MEA and
MKK groups for providing the necessary infrastructure and support
during the assembly, commissioning, operation, and disassembly of the
experiment. The research and engineering group from MIT-Bates was
invaluable in all phases of the experiment, from disassembling BLAST
and shipping components to DESY to overcoming numerous unanticipated
problems during the installation and operation of the experiment,
particularly with the target and vacuum systems.

We would like to thank E.\@ Steffens for numerous suggestions and
helpful discussions during the initial development of the experiment.

Finally, we gratefully acknowledge the DESY directorate, particularly
Prof.\@ Heuer and Prof.\@ Mnich, and the DESY Physics Research
Committee for their support, advice, and encouragement from the start
of the proposal.

This work was supported by numerous funding agencies which we
gratefully acknowledge: the Ministry of Education and Science of
Armenia, the Deutsche Forschungsgemeinschaft , the European
Community-Research Infrastructure Activity , the United Kingdom
Science and Technology Facilities Council and the Scottish
Universities Physics Alliance, the United States Department of Energy
and the National Science Foundation, and the Ministry of Education and
Science of the Russian Federation. R. Milner also acknowledges the
generous support of the Alexander von Humboldt Foundation, Germany.

\bibliographystyle{elsarticle-num.bst}
\bibliography{OLYMPUS}

\end{document}